\documentclass[%
 aip, jcp,
 amsmath,amssymb,amsfonts,
 reprint,%
 superscriptaddress
]{revtex4-1}

\usepackage[dvips]{graphicx}
\usepackage{dcolumn}
\usepackage{bm}
\usepackage{amsmath}
\usepackage{latexsym}
\usepackage{color}

\usepackage[utf8]{inputenc}
\usepackage[T1]{fontenc}
\usepackage{mathptmx}
\usepackage{hyperref}
\hypersetup{
   bookmarks=true,         
   colorlinks=true,       
   linkcolor=magenta,          
   citecolor=magenta,        
   filecolor=magenta,      
   urlcolor=blue           
}

\begin{document}


\title{A bosonic perspective on the classical mapping of fermionic quantum dynamics}

\author{Jing Sun} 
 \email{ny182@uni-heidelberg.de}
\author{Sudip Sasmal}%
 \email{sudip.sasmal@pci.uni-heidelberg.de}
\author{Oriol Vendrell}%
 \email{oriol.vendrell@uni-heidelberg.de}
\affiliation{
    Theoretische Chemie, Physikalisch-Chemisches Institut, Universität
    Heidelberg, Germany.
}

\date{\today} 

\begin{abstract}

We consider the application of the original Meyer-Miller (MM) Hamiltonian to mapping fermionic
quantum dynamics to classical equations of motion.
Non-interacting fermionic and bosonic systems share the same one-body density dynamics
when evolving from the same initial
many-body state. The MM classical mapping is exact for non-interacting bosons, and therefore
it yields the exact time-dependent one-body density for non-interacting fermions as well.
Starting from this observation, the MM mapping is compared to different mappings
specific for fermionic systems, namely the spin mapping (SM) with
and without including a Jordan-Wigner transformation, and the Li-Miller mapping (LMM).
For non-interacting systems, the inclusion of fermionic anti-symmetry through the Jordan-Wigner
transform does not lead to any improvement in the performance of the mappings and instead it
worsens the classical description.
For an interacting impurity model and for models of excitonic energy transfer, the MM and LMM mappings perform
similarly, and in some cases the former outperforms the latter when compared to a full quantum description.
The classical mappings are able to capture
interference effects, both constructive and destructive, that originate from equivalent energy transfer pathways
in the models.
\end{abstract}

\maketitle

\section{Introduction}

The pursuit of classical and semi-classical theories to approximate the quantum
dynamics of molecular systems has a long
history,~\cite{Miller1986,mey79:3214,Meyer1979_2,mey80:2272,Heller1981,%
tul90:1061,ham94:4657,Martens1997,%
Kapral1999, Kernan2008, Stock1997,
Thoss1999,%
Ben2000, Curchod2018, Tavernelli2013, Curchod2011, Agostini2016, Runeson2019,Runeson2020, lan21:024111}
and has been motivated by their cost-effectiveness and the fact that chemical
dynamics take place often in an energy and density-of-states regime where a
classical description can be meaningful.
The most wide-spread type of approaches describes the nuclear-electronic
non-adiabatic dynamics in molecules by splitting the degrees of freedom into
nuclear and electronic subspaces. The nuclei are then set to evolve classically
in a potential derived from the interaction with the electrons, while
quantum mechanics is maintained for the electronic subspace.
The latter dynamics can be described by a time-dependent Schrödinger equation in
the discrete space of diabatic or adiabatic electronic states, or by mapping
this discrete space into a set of classical variables, which then evolve
together with the nuclei under an overall classical Hamiltonian. Several mappings
exist for the electronic degrees of freedom, for example the original
Meyer-Miller (MM) Hamiltonian consisting of one harmonic oscillator per
electronic state, and different flavors of the spin-mapping (SM) Hamiltonian,
which use classical spin degrees of freedom to map the state of the electronic
subspace~\cite{mey80:2272,Cotton2015}.
Importantly, Stock and Thoss showed that the nuclear-electronic MM Hamiltonian
yields the exact quantum dynamics once re-quantized,~\cite{Stock1997}
and this mapping is also known as MMST.%

Although the equations of motion (EOM) of mapping approaches are fully
classical, this fact is, by itself, not necessarily an
approximation. The classical EOM of the MM mapping for the electronic states are
formally equivalent to solving the Schrödinger equation
they have a close formal resemblance with the Ehrenfest
method~\cite{mey79:3214,mey80:2272}. The differences are, however, significant and related
to how initial conditions are sampled and to how observables are derived from
the trajectories~\cite{mey79:3214,mey80:2272,Cotton2013,cot15:12138}.
One can argue that mapping approaches provide a more fundamental answer to
the question of how to
mix classical and quantum degrees of freedom than, e.g.  Ehrenfest 
or surface-hopping approaches
~\cite{tul90:1061,ham94:4657}, and often outperform them in benchmark
applications~\cite{Runeson2019,Runeson2020,Liang2018,Gao2020}.

Miller and White took the pioneering step to extend the concept of classical
mappings to the treatment of electrons (fermions) under a second-quantized
Hamiltonian~\cite{Miller1986}. This step can be motivated by the formal
similarity between the second-quantized Hamiltonian for bosons and fermions, and
the fact that the bosonic creation/annihilation operators are analogous to the
ladder operators of interacting harmonic oscillators, which can subsequently be
downgraded to classical variables to obtain a useful mapping.
In their work, Miller and White arrived at a mapping for the fermionic operators
using the Heisenberg correspondence relation~\cite{Miller1976,Miller1978} between matrix elements and
classical variables. By construction, the Miller-White (MW) mapping
respects the sign-change rules of the
the commutation relations of fermionic operators.
Although it is not
presented in this way in Ref.~\citenum{Miller1986}, this mapping can be
alternatively reached by first performing a (exact) Jordan-Wigner transformation of the
fermionic Hamiltonian into a corresponding chain of spins~\cite{jor28:631},
where so-called sign-change operators $(1-2\hat{n}_j)$ appear, and then taking a
spin-mapping (SM)~\cite{mey79:3214,cot15:12138,Meyer1979_2,mey80:2272} for each fermionic degree of freedom (cf. Appendix).

In their original work, the MW mapping was not applied to dynamical processes but
it was demonstrated that it yields the correct energy for
selected electronic configurations of the helium atom and of the hydrogen
molecule.
Subsequent works reformulated the MW
mapping on the basis of a Cartesian Hamiltonian, like in the original MM mapping, while
still preserving the sign-change rules
of the fermionic operators.\cite{Li2012,Li2013,Li2014,Levy2019} This mapping, called Li-Miller mapping (LMM)
in a recent publication~\cite{Levy2019}, yields the exact dynamics of the
fermions in Fock space for non-interacting Hamiltonians~\cite{Li2012}.
Based on these developments, promising results have been obtained in studies of
electronic dynamics and quantum transport in molecular junctions and quantum
dots.~\cite{Zwanziger1986, Voorhis2004, Swenson2011,Cotton2013,cot15:12138,Li2012,Li2013,Li2014,Liu2016,Levy2019}
These results indicate that a broader range of electronic
processes in molecules may be approachable through classical mappings of the
electronic degrees of freedom in Fock space, which is one of the
motivations for this work.



It turns out that the original MM mapping applied to non-interacting fermionic
Hamiltonians, in the same way as the LMM, delivers the exact dynamics of
the system when the initial state is a physical fermionic state (cf.
Sec.~\ref{section:theory} below).
Starting from this observation, in this study we explore this formal connection
and examine the applicability of the original MM mapping to the quantum dynamics
of electrons in closed systems in a second quantized setting.
In section~\ref{section:theory}, the formal connection that explains this
equivalence for non-interacting systems is laid down.  We proceed by
describing the relation between the initial phase space density of the
classically mapped system and the initial configuration of the electrons, and
propose strategies to sample this phase space density.
Section~\ref{section:result} compares the MM mapping with exact quantum results
and with different mappings explicitly designed for fermions, namely the SM with
and without inclusion of antisymmetry (the latter corresponds to the original MW
mapping), and to the LMM.
We compare Hubbard and impurity Hamiltonians, with and without interactions, and
consider as well a model for excitonic energy transfer between chromophores. In
this model with interactions we show that the classical MM mapping is able to
capture interference effects caused by the presence of different energy
transfer pathways leading to the same final state, both when the interferences
are constructive and destructive.
For the model Hamiltonians and parameter ranges
considered, we show how the MM mapping performs comparably to the LMM, even
outperforming it in some cases, although it uses half the number of classical
variables to map the state of the electrons. MM invariably outperforms the SM mapping
with (equivalent to MW) and without inclusion of anti-symmetry.

\section{Theory}
\label{section:theory}

\subsection{Equation of motion concerning bosons and fermions}

The second-quantized Hamiltonian for a many-body system of fermions (electrons
in our case) and bosons takes the general form
\begin{align}
    \label{eq:Ham}
    \hat{H} & = \hat{H}^{(1)} + \hat{H}^{(2)} \\
    \hat{H} & =\sum_{ij}^F h_{ij}a^{\dagger}_{i}a_{j}+
    \frac{1}{2}\sum_{ijkl}^FV_{ijkl}a^{\dagger}_{i}a^{\dagger}_{j}a_{l}a_{k} ,
\end{align}
where i, j, k and l run over all single particle states and $F$ is the number
of single-particle basis functions used to expand the Fock space.
In this paper we will refer to the annihilation (creation) operators as $\hat{a}_j^{(\dagger)}$ in
general and specialize them to $\hat{b}_j$ for bosons and $\hat{c}_j$ for
fermions whenever this distinction is needed. These operators
obey the respective commutation and (anti-)commutation relations
\begin{align}
    \label{eq:commutation}
    [\hat{b}_i, \hat{b}_j^\dagger] = \delta_{ij};\quad &
    [\hat{b}_i, \hat{b}_j] = [\hat{b}_i^\dagger, \hat{b}_j^\dagger] = 0; \\
    \{\hat{c}_i, \hat{c}_j^\dagger\} = \delta_{ij};\quad &
    \{\hat{c}_i, \hat{c}_j\} = \{\hat{c}_i^\dagger, \hat{c}_j^\dagger\} = 0.
\end{align}

As is well known, for bosons one can identify the creation and annihilation
operators with the ladder operators of a set of harmonic oscillators, one for
each bosonic mode, which obey the same commutation relations. The ladder
operators can be expressed using the corresponding
positions and momenta
\begin{align}
\begin{split}
    \label{eq:b-to-qp}
    \hat{b}_j & \mapsto \frac{1}{\sqrt{2}} (i\hat{p}_j + \hat{q}_j)\\
    \hat{b}_j^\dagger & \mapsto \frac{1}{\sqrt{2}} (-i\hat{p}_j + \hat{q}_j),
\end{split}
\end{align}
leading to a form of the bosonic Hamiltonian with a simple classical analog. Note that $\hbar$ is set to be $1$ which is also applied to all the other equations in below. The equations of motion of the classical positions and momenta follow from the usual prescription: substitute the commutator with the Poisson bracket in the
corresponding Heisenberg equations of motion, or equivalently derive Hamilton's
equations directly from the classical form of the Hamiltonian.
Already Miller and White in their seminal paper recognized that, would fermions
obey the same commutation relations as bosons, this analogy would
offer a straightforward path towards a classical mapping for fermions~\cite{Miller1986}.

It is useful to first discuss in some detail the properties of a classical
approximation to the dynamics of the $\hat{H}^{(1)}$ term for bosons before considering fermions.
Using the
relations~(\ref{eq:b-to-qp}) and replacing quantum operators with classical
variables one arrives at the Hamiltonian function
\begin{align}
    \label{eq:h1cl}
    {H}^{(1)}_{\text{B;cl}}(\mathbf{p}, \mathbf{q}) =
    \frac{1}{2} \left( \mathbf{p}^+ \, \mathbf{h}\, \mathbf{p} + 
                       \mathbf{q}^+ \, \mathbf{h}\, \mathbf{q} \right)
                    - \gamma \; \mathrm{Tr}[\mathbf{h}],
\end{align}
where the column arrays $\mathbf{p}$ and $\mathbf{q}$ collect all momenta and
positions, respectively, and $\mathbf{h}$ is the matrix with elements $h_{ij}$.
The $\gamma$ factor takes values between 0 and $1/2$ and it does not affect the
classical equations of motion. However, it determines the amount of zero-point
energy available to the classical system when sampling initial
conditions~\cite{mey79:3214,cot16:983}.
 Hence, the results obtained with the mapping for interacting systems depend on
 $\gamma$. In the numerical results discussed below we have made the experience,
 similarly to other applications in the literature~\cite{Cotton2013,cot15:12138},
 that $\gamma\approx 0.366$ works best, and this is the value we keep throughout.
This Hamiltonian is identical with the electronic part of the original MM
Hamiltonian~\cite{mey79:3214,Stock1997,Cotton2013}, although in this case the
classical oscillators do not map the amplitudes of the electronic states, but instead the
number of particles in each single particle state, $n_j = (p_j^2 + q_j^2)/2 -
\gamma$.
%

%
The linear Hamilton's equations arising from Hamiltonian~(\ref{eq:h1cl}) are
\begin{align}
    \label{eq:hamilton1}
    \overset{\bullet}{\mathbf{q}} = \mathbf{h}\, \mathbf{p};\quad
   -\overset{\bullet}{\mathbf{p}} = \mathbf{h}\, \mathbf{q}
\end{align}
and they are equivalent to the time-dependent Schrödinger equation (TDSE)
\begin{align}
    \label{eq:tdseB}
    i \overset{\bullet}{\mathbf{c}} = \mathbf{h} \, \mathbf{c},
\end{align}
when the complex coefficients
$\mathbf{c} = (\mathbf{q} + i \mathbf{p})/\sqrt{2}$
are introduced.
This analogy between Hamilton's equations and the TDSE was already recognised by
Meyer and Miller in their original work on the MM
mapping~\cite{mey79:3214,Cotton2013}.
However, in the MM case, the coefficients in Eq.~(\ref{eq:tdseB}) are normalized
to 1, whereas in the non-interacting bosonic case they are normalized to the
number of particles, $P = \mathbf{c}^* \mathbf{c}$.
This results in the caveat that unique Fock-space states
$|\mathbf{n}\rangle\equiv|\{n_1,\ldots,n_F\}\rangle$ do not have a unique
parametrization in terms of $\mathbf{c}$ (or $\mathbf{q}$, $\mathbf{p}$) because
an arbitrary phase can be added to every element of $\mathbf{c}$ without
altering $|\mathbf{n}\rangle$, whereby different parametrizations of
$|\mathbf{n}\rangle$ lead to different dynamics of the expansion coefficients.
%
This fact is illustrated numerically in some of the examples in
Section~\ref{section:result}.
Remacle and Levine had already encountered the non-uniqueness of
Eq.~(\ref{eq:hamilton1}, \ref{eq:tdseB}) to describe a \emph{many-body} system
of non-interacting electrons through a \emph{single classical trajectory}, but a
workaround was not proposed and only applications to one-electron systems were
discussed~\cite{Remacle2000}.

This caveat can be resolved, still for non-interacting bosons, by 
considering the time-evolution of the corresponding
Wigner density function~\cite{tannor-book, imr67:1097}
\begin{align}
    \label{eq:wigner}
    \overset{\bullet}{\rho}_W(\mathbf{q},\mathbf{p}) =
    -2 {H}^{(1)}(\mathbf{q},\mathbf{p})
    \sin\left( \frac{ \Lambda }{2}\right) \rho_W(\mathbf{q},\mathbf{p}),
\end{align}
where $\Lambda=
      \overset{\leftharpoonup}{\nabla_q} \overset{\rightharpoonup}{\nabla_p} -
      \overset{\leftharpoonup}{\nabla_p} \overset{\rightharpoonup}{\nabla_q}$,
and $A_W \Lambda B_W\equiv\{A_W, B_W\}$ indicates the
Poisson bracket.
Under the quadratic Hamiltonian $H^{(1)}$ only the lowest order expansion of
the sine function in Eq.~(\ref{eq:wigner}) contributes and one immediately
arrives at the classical Liouville equation for the density,
\begin{align}
    \label{eq:liouville}
    \overset{\bullet}{\rho}(\mathbf{q},\mathbf{p}) =
    \{ {H}^{(1)}(\mathbf{q},\mathbf{p}), \rho(\mathbf{q},\mathbf{p})
    \},
\end{align}
which delivers the
\emph{exact} phase-space dynamics of the non-interacting system.
This density can be propagated as a swarm of classical trajectories
matching the quantum mechanical initial conditions,
which is the same as the Wigner classical approximation
of Heller~\cite{Heller1981}. For non-interacting bosons this is
not an approximation, but an alternative way to compute the exact dynamics of the
system.

Up to now our discussion has been centered on bosonic particles and we have not
made progress yet towards our primary goal, the classical mapping of
fermionic particles.
Let us return to the quantum mechanical problem and consider the time evolution
of the single-particle density matrix elements
$\langle \hat{a}_j^\dagger \hat{a}_i\rangle$
\begin{align}
    \label{eq:pops}
    \frac{d}{dt} \langle \hat{a}_j^\dagger \hat{a}_i\rangle =
    i
    \left(
        \sum_k^F h_{kj} \langle \hat{a}_k^\dagger \hat{a}_i \rangle
        -\sum_l^F h_{il}  \langle \hat{a}_j^\dagger \hat{a}_l \rangle
    \right).
\end{align}
Equation~(\ref{eq:pops}) is a closed expression: once the $F$ populations
$\langle \hat{n}_j \rangle$ and the $F(F-1)$ off-diagonal terms $\langle \hat{a}_j^\dagger \hat{a}_i \rangle$
are specified, their evolution follows uniquely.
They key observation is that Eq.~(\ref{eq:pops}) is identical for bosons and for
fermions. In other words, the same initial one-body density of non-interacting
bosonic and fermionic systems has the same time evolution.

The important consequence for our purposes is that the original MM mapping in
Eq.~(\ref{eq:h1cl}) also delivers the exact quantum dynamics of
non-interacting Hamiltonians for many-body fermionic systems
\emph{once
the initial one-body density matrix of the fermionic state has been mapped onto the
corresponding phase-space density in Eq.~(\ref{eq:liouville})}.
This result will be illustrated with various numerical examples below.
Note that we have made no attempt to include the anti-symmetry of the fermions
through a Jordan-Wigner transformation,~\cite{jor28:631} nor
to limit the maximum occupation of the orbitals with a spin mapping (SM)
model~\cite{Miller1986,cot15:12138}.
As is known~\cite{Li2012,Li2013,Li2014}, the mappings based on those concepts do not deliver the
\emph{exact} dynamics in the non-interacting limit and, as we will show, they indeed perform poorly
compared to the MM mapping, with and without interactions.

The exact dynamics in the non-interacting limit is reproduced as well by the LMM
mapping, which is explicitly devised for
fermions~\cite{Levy2019,Li2012,Li2013,Li2014}, but at the cost of doubling the
size of the classical phase-space compared to the MM mapping. For the range of
examples with interactions considered in this work, though, we could not see any
significant advantage of the LMM over the MM mapping.
It is worth mentioning, however, that the LMM has been developed for, and
applied to, semiclassical initial value
calculations,\cite{Li2012,Li2013,Li2014} whereas in this work we
consider it in a fully linearized, classical context.
%
%
For completeness, both the SM and LMM are described in the Appendix.

Finally, two equivalent propagation strategies are available for non-interacting
systems:
(i) The initial fermionic Fock-space state is mapped onto a corresponding
phase-space distribution, Eq.~(\ref{eq:liouville}), which can then be
conveniently discretized as $N$ phase-space trajectories, each evolving as
$2F$ coupled Hamilton's equations.
(ii) The one-body density matrix corresponding to the initial fermionic state is
propagated according to the $F^2$ coupled differential Eqs.~(\ref{eq:pops}).
The former strategy is our working approximation for
interacting Hamiltonians. The latter strategy,
propagating the one-body density matrix, becomes essentially
equivalent to solving the full quantum-mechanical problem once interactions are
included, and we do not consider approximations along that line.

\subsection{Sampling of initial conditions}

Using the relations~(\ref{eq:b-to-qp}) with the single-particle density matrix
elements $\langle \hat{a}_j^\dagger \hat{a}_i\rangle$ and equating
expectation values to classical phase-space averages,
one can write the relation
 \begin{align}
    \label{eq:nij_pq_int}
    \langle \hat{a}^{\dagger}_j\hat{a}_j\rangle & =
     \frac{1}{2}\int
    \left( \right.
     p_{j}^2 +q_{j}^2 - 2\gamma
    \left .\right) \rho(\mathbf{q},\mathbf{p})\;\;
    d^F\!\!\!q\;\; d^F\!\!\!p \\\nonumber
    \langle \hat{a}^{\dagger}_j\hat{a}_i\rangle & =
     \frac{1}{2}\int
    \left( \right.
     p_{j}p_{i}
    +q_{j}q_{i}
     -ip_{j}q_{i}
    +iq_{j}p_{i}
    \left .\right) \rho(\mathbf{q},\mathbf{p})\;\;
    d^F\!\!\!q\;\; d^F\!\!\!p
 \end{align}
between the quantum mechanical matrix element of the one-body density
and the classical phase-space average of position and momenta
for the MM mapping.
The phase-space density can be represented by a
discrete set of $N$ phase-space points $(\mathbf{p}^{(k)},
\mathbf{q}^{(k)})$,
 \begin{align}
    \label{eq:nij_pq}
    \langle \hat{a}^{\dagger}_j\hat{a}_j\rangle & = \frac{1}{{N}}
    \sum_{k=1}^{{N}}\frac{1}{2}
    \left( \right.
     {p_{j}^{(k)}}^2 + {q_{j}^{(k)}}^2 - 2\gamma
    \left .\right) \\\nonumber
    \langle \hat{a}^{\dagger}_j\hat{a}_i\rangle & = \frac{1}{{N}}
    \sum_{k=1}^{{N}}\frac{1}{2}
    \left( \right.
     p_{j}^{(k)}p_{i}^{(k)}
    +q_{j}^{(k)}q_{i}^{(k)}
     -ip_{j}^{(k)}q_{i}^{(k)}
    +iq_{j}^{(k)}p_{i}^{(k)}
    \left .\right).
 \end{align}
%
It is more convenient to discuss the sampling of initial conditions in terms of
the corresponding action-angle variables,~\cite{mey79:3214,cot16:983}
whose relation to $(q_j,p_j)$ follows compactly as
\begin{align}
    \label{eq:actang}
     q_j + i p_j = \sqrt{2(n_j+\gamma)} e^{i\phi_j}.
\end{align}
Using Eq.~(\ref{eq:actang}), Relation~(\ref{eq:nij_pq}) becomes
\begin{subequations}
 \label{eq:nij_actang}
 \begin{align}
    \langle \hat{a}^{\dagger}_j\hat{a}_j\rangle & = \frac{1}{{N}}
    \sum_{k=1}^{{N}} n_j^{(k)} \\
    \langle \hat{a}^{\dagger}_j\hat{a}_i\rangle & = \frac{1}{{N}}
    \sum_{k=1}^{{N}}
     \sqrt{(n_j^{(k)}+\gamma)( n_i^{(k)}+\gamma)} e^{i(\phi_i^{(k)} - \phi_j^{(k)})}.
 \end{align}
\end{subequations}

Let us consider first the simple case that the initial state in Fock space
corresponds to a single configuration with one single electron,
$|j\rangle\equiv |0_1,\ldots,1_j,\ldots 0_F\rangle$,
in the occupation number representation.
In this case $\langle \hat{a}_j^\dagger \hat{a}_j\rangle=1$ and all other matrix
elements of the one-body density matrix are zero.
Therefore the one-body density matrix can be mapped to a single phase-space
point with action variables $n_j=1$, $n_l=0$ for $l\neq j$ and all angle
variables $\phi_m=0$. Integrating the corresponding classical trajectory is of
course equivalent to solving the corresponding TDSE, Eq.~(\ref{eq:tdseB}).
Another illustrative example is the single-electron case where the initial state is a
linear superposition of configurations
\begin{align}
    \label{eq:superposition}
    |\Psi\rangle = \sum_{j=1}^F C_j |j\rangle
\end{align}
where $C_j$ are complex expansion coefficients and $|j\rangle$ are the
single-electron configurations defined
above.
Now, by setting $\gamma=0$, Eqs.~(\ref{eq:nij_actang}) can be fulfilled
simultaneously
by a single phase-space point where the action-angle variables
are chosen such that $C_j = \sqrt{n_j} e^{i\phi_j}$.
This situation is analogous to setting $\gamma=0$
in the original MM mapping for electronic states, which then reverts to an
Ehrenfest model with a single trajectory.
If $\gamma\neq0$, the one-body density of the single-electron state
(\ref{eq:superposition}) cannot be described, in general, by a single
phase-space point according to Eqs.~(\ref{eq:nij_actang}).

A more useful case corresponds to a single-configuration many-body state of the
form
$|1_1,1_2,\ldots,1_m,0_l,\ldots,0_F\rangle$,
for example a Hartree-Fock approximation of the ground electronic state or an
excited state that can be initially well described as a single configuration.
For such a state, the one-body density matrix is diagonal with
$\langle \hat{a}_j^\dagger \hat{a}_j\rangle=n_j$. Independently of the choice of
$\gamma$, it is clearly impossible to fulfill, e.g., that
\begin{subequations}
\begin{align}
    \langle \hat{a}_1^\dagger \hat{a}_1\rangle & = \langle \hat{a}_2^\dagger \hat{a}_2\rangle = 1 \\
    \langle \hat{a}_2^\dagger \hat{a}_1\rangle & = 0
\end{align}
\end{subequations}
with one single phase-space point.
One can think of different ways to sample initial conditions that fulfill
Eqs.~(\ref{eq:nij_actang}), for example by introducing a specific number
$N$ of initial phase-space points and then solving Eqs.~(\ref{eq:nij_actang}) for
all unknown $n_j^{(k)}, \phi_j^{(k)}$ parameters.
For instance, it is possible to set all $n_j^{(k)}$ of each $k$-point
equal to the corresponding spin-orbital populations,
Eq.~(\ref{eq:nij_actang}a), while the remaining angle variables of each
$k$-point must then be determined such that the remaining sum of phases in
Eq.~(\ref{eq:nij_actang}b) vanishes.
This corresponds to determining $F\times N$ unknowns for the corresponding set
of $F(F-1)/2$ non-linear equations and it is not clear, \emph{a priori}, what is the smallest
number $N$ of phase-space points needed to map a specific density.
We have made no attempt to systematically solve Eq.~(\ref{eq:nij_actang}b) for
the angles of general initial states. We illustrate the concept with
a simple example with $F=3$ spin-orbitals and 2 particles,
which will be considered later numerically.
Starting with state $|1,1,0\rangle$,  one needs at least $N=4$ points
(trajectories) to reproduce the corresponding one-body density matrix and one can easily
verify that the angles in Table~\ref{Tab:DS_ex} result in the cancellation of the
phase factors in Eq.~(\ref{eq:nij_actang}b).
Clearly, the choice of angles in Table~\ref{Tab:DS_ex} is not unique,
only the angle differences between each DOF matter.
\begin{table}
\begin{tabular}{c c c c}
  & $\phi_1$ & $\phi_2$ & $\phi_3$ \\
 \hline
 \hline
 Trj 1 & 0 & $\pi$/4 & 3$\pi$/4 \\ 
 Trj 2 & 0 & 3$\pi$/4 & $\pi$/4 \\
 Trj 3 & 0 & 5$\pi$/4 & 7$\pi$/4 \\
 Trj 4 & 0 &7$\pi$/4 & 5$\pi$/4 \\
 \hline
\end{tabular}
\caption{\label{Tab:DS_ex}In a system with 3 DOFs, an instance is presented by solving Equation~(\ref{eq:nij_pq}).}
\end{table}

Alternatively to the discrete sampling just introduced, it is possible to
perform a random sampling of the initial angle variables.
For single-configuration initial states, where all off-diagonal matrix elements
of the one-body density matrix are zero, this is the most straightforward way to
determine initial conditions. The disadvantage is that one may end up
propagating more classical trajectories than strictly needed.
For systems with many single-particle states (orbitals) $F$, this is however a straightforward
way to proceed, as compared to solving Eq.~(\ref{eq:nij_actang}b) for some
specific discretization $N$.
In Section~\ref{sec:non-inter} we numerically illustrate how, for non-interacting
systems, both a small set of trajectories with
initial angles chosen to fulfill Eq.~(\ref{eq:nij_actang}b), and a randomly
sampled ensemble with a larger number of trajectories, yield the exact
one-body dynamics of the system.
For interacting Hamiltonians, Eq.~(\ref{eq:liouville}) is not equivalent
to Eq.~(\ref{eq:wigner}) and therefore the former does not reproduce the exact dynamics
of the one-body density. Moreover, different classical phase-space distributions
representing the same initial one-body density result in different time
evolutions.

Finally, multi-configurational many-body states
have, in general, a non-diagonal one-body density matrix that cannot be
factorized. Because the off-diagonal elements are not necessarily equal to zero,
the random sampling strategy cannot be applied, and one
is left with the alternative of solving Eqs.~(\ref{eq:nij_actang}) to determine the
initial ensemble of phase space points. In this work we focus instead on
single-configuration initial states.

\section{Results and discussion}
\label{section:result}

In the following, we compare classical mappings with exact quantum
calculations in both non-interacting and interacting systems. 
Except in the discussion of Fig.~\ref{fig:3site12_RS}a, where an explicit comparison between
discrete and random sampling is made, the initial conditions for all classical
mapping calculations and for all benchmarked mappings are generated by random
sampling of the angle variables. All quantum calculations are obtained using the
multi-configuration time-dependent Hartree (MCTDH)
approach~\cite{Meyer1990,bec00:1,mctdh:MLpackage} in the second-quantization
representation (SQR)
formulation~\cite{wan09:024114,man17:064117,sas2020:154110}, which is equally applicable to
fermions and bosons.
In Section~\ref{sec:non-inter}, we address several many-body systems without
interactions, where the performance of the different mappings and the
inclusion of anti-symmetry can be compared and discussed.
Subsequently, Section~\ref{sec:inter} concerns with comparisons between
classical mappings and exact quantum results for several many-body systems with
interactions, ranging from Hubbard-like Hamiltonians to a
model for excitonic energy transfer.

\subsection{Non-interacting systems} \label{sec:non-inter}

\subsubsection{2- and 3-site systems}

\begin{figure}
    \includegraphics[width=0.95\columnwidth]{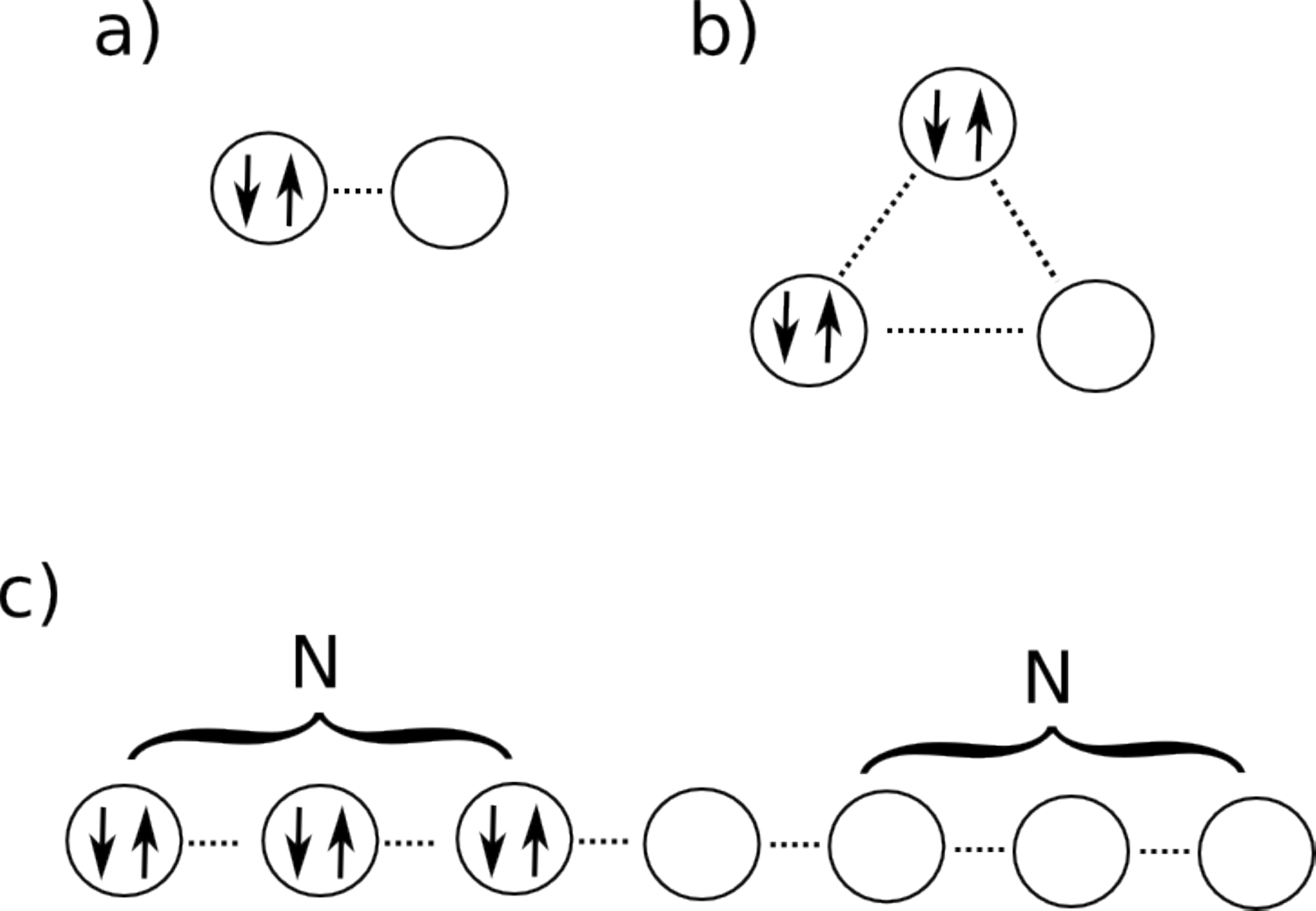}
    \caption{\label{fig:topo_site}
    Structure of a) 2-site and b) 3-site tight-binding systems.
    The arrows indicate the occupation of the orbitals at $t=0$ and the
    dashed lines represent the one-body transfer integrals $h_{ij}$.}
\end{figure}

First of all, we compare the MM and SM mapping with the exact electron dynamics
in 2- and 3-site tight-binding Hamiltonians
\begin{align}
    \label{eq:tightB}
    \hat{H}_\text{tb} = \sum_{\langle i,j\rangle}^F
                        \sum_{\sigma=\alpha,\beta}
    T(\hat{c}^\dagger_{j,\sigma} \hat{c}_{i,\sigma} + \text{h.c}),
\end{align}
where $\langle i,j\rangle$ indicates that the sum runs over nearest neighbors
only.
A diagram indicating the interactions between the sites is shown in
Fig.~\ref{fig:topo_site}, where orbital interaction terms are marked
with a dotted line and are all set to $T=-0.05$~Hartree.
For the MM mapping, one simply applies the relations (\ref{eq:b-to-qp}) to
Hamiltonian~(\ref{eq:tightB}) to reach the classical Hamiltonian function.  For
the SM mapping, one can introduce the anti-symmetry of
the fermionic quantum-mechanical operators by first transforming the Hamiltonian
to an equivalent form based on spin-$1/2$ degrees of freedom, the so-called Jordan-Wigner
transformation~\cite{Jordan1993} (JWT),
\begin{subequations}
\label{eq:jorwig}
\begin{eqnarray}
\hat{c}_i^{\dagger} \to \prod_{k=1}^{i-1} \hat{S}_k \cdot \hat{\sigma}_i^{+}\\
\hat{c}_i \to \prod_{k=1}^{i-1} \hat{S}_k \cdot \hat{\sigma}_i^{-}
\end{eqnarray}
\end{subequations}
with $S_k=\exp(i\pi\hat{n}_k)$. This transformation is also the
key ingredient involved in the description of fermions in second-quantization within
the MCTDH method.~\cite{wan09:024114,sas2020:154110}
Here, $\hat{\sigma}_i^{\pm}$ are spin-$1/2$ ladder operators and $\hat{S}_k$
are sign-change operators acting locally on index $k$ such that
$\hat{S}_k|0_k\rangle=|0_k\rangle;\;\;\hat{S}_k|1_k\rangle=-|1_k\rangle$, and the spin states are
used to indicate occupation:
$|\!\!\downarrow_k\rangle\to|0_k\rangle$;
$|\!\!\uparrow_k\rangle\to|1_k\rangle$.
The sign-change operators can also be written as $\hat{S}_k=1-2\hat{n}_k$, where
$\hat{n_k}= \hat{\sigma}_k^{+} \hat{\sigma}_k^{-}$. This substitution together
with the JWT relations~(\ref{eq:jorwig}) and the SM result precisely in the
original MW mapping for fermionic Hamiltonians
(cf. Eqs. 2.10 and 2.13 in Ref.~\citenum{Miller1986}).

\begin{figure}
\includegraphics[width=0.95\columnwidth]{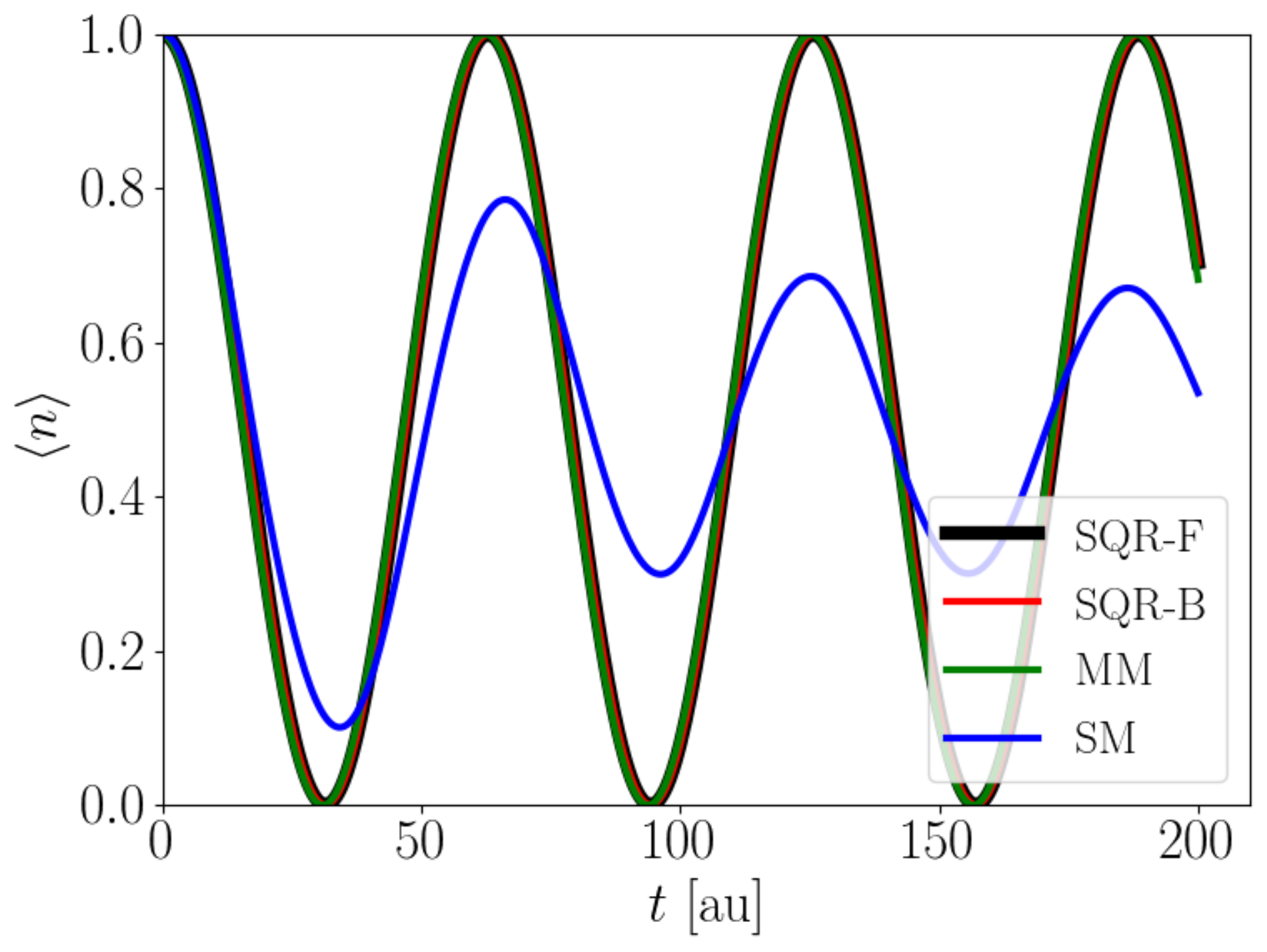}
\caption{\label{fig:2site} Time-dependent population of site 1, $n_{1\alpha}$,
in the 2-sites tight-binding system for exact fermionic and bosonic dynamics, and
for the MM and SM mapping approaches.
The transfer integral is set to $T=-0.05$ Hartree.}
\end{figure}

Applying the JWT to the tight-binding binding Hamiltonian~(\ref{eq:tightB}) with
linear topology one arrives at a pure spin-$1/2$ Hamiltonian where all $\hat{S}_k$ operators have cancelled. Therefore, SM and MW (meaning JWT+SM) are
equivalent.
We examine now the population dynamics of the 2-site tight-binding model
starting with site 1 fully populated, Fig.~\ref{fig:topo_site}a.
Since we are dealing with a quantum spin Hamiltonian, one would now be inclined
to think that a classical SM representation should deliver a more accurate
approximation to the quantum dynamics of the model than MM. We know that this cannot be
true because, as already discussed, the MM representation yields the exact
dynamics in the non-interacting case. Fig~\ref{fig:2site} confirms this and also
shows how the SM fails to reproduce the exact population dynamics.
In fact, the equations of motion of the SM model are not fully linear, which
leads to continuous dissipation of the classical trajectories in phase space,
and hence to their failure to reproduce the correct amplitude of the population oscillations.
This observation is not new, and the reason why Miller and collaborators have
developed the LMM Cartesian (oscillator-based) version of the original MW
mapping in recent years~\cite{Levy2019,Li2012,Li2013,Li2014}.

\begin{figure}
    \includegraphics[width=0.95\columnwidth]{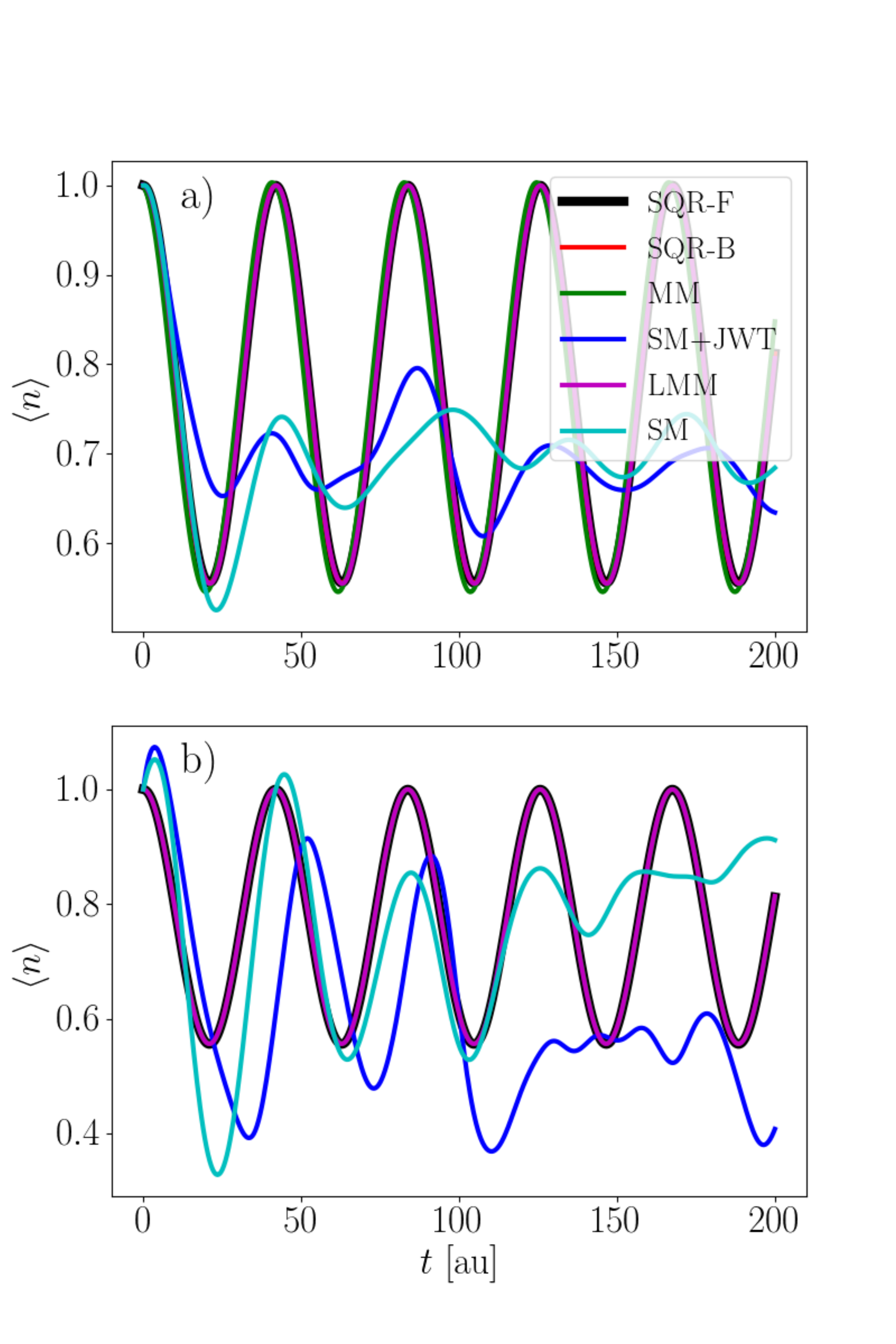}
    \caption{\label{fig:3site12_RS}
    Time-dependent population of site 1, $n_{1\alpha}$,
    in the 3-sites cyclic tight-binding system for exact
    fermionic and bosonic dynamics, and for the MM, SM, SM+JWT and LMM mapping approaches.
    (a) initial conditions generated by homogeneous random
    sampling. (b) discrete sampling initial conditions with $N=8$ trajectories.
    In both cases all approaches yield the exact result, except the
    SM and SM+JWT ones.
    }
\end{figure}

Matters turn more interesting when considering a 3-site tight-binding
Hamiltonian with cyclic topology, where the initial occupation of the sites is
shown in Fig.~\ref{fig:topo_site}b.
Now, the sign-change operator $\hat{S}_2$ survives in the term proportional to
$h_{13}$, meaning that the SM and SM+JWT mappings result in different classical
Hamiltonians.
As expected from our theoretical considerations, the MM mapping reproduces the
exact population dynamics, even if now the JWT yields a sign-change operator.
The same is found to be true for the LMM
(cf. Fig.~\ref{fig:3site12_RS}a), which explicitly considers the
anti-commutation relations of the fermionic operators.
The SM+JWT mapping, instead, again fails to represent the correct
site-population dynamics, now even more dramatically than before.

For comparison, we also examine the dynamics of the same 3-site Hamiltonian(6 DOFs) but
now the initial one-body density matrix is mapped to $N=8$ trajectories. 
%
The population dynamics in Fig.~\ref{fig:3site12_RS}b are seen to be exact for
the MM mapping and LMM, whereas they fail to reproduce the quantum result for
both the SM and SM+JWT mappings. The two SM-based mappings yield different
results when compared among themselves and the SM mapping follows more closely
the oscillations of the quantum mechanical population dynamics.
From this observation we obtain a first hint at the fact that, including
antisymmetry with a JWT before the performing the classical mapping, which is
implicit in the MW and LMM treatments~\cite{Miller1986,Levy2019}, may not
actually be relevant to the mapping of fermions and may indeed be
counterproductive, resulting in a classical Hamiltonian with higher-order
interactions than simpler mappings and a more chaotic classical dynamics that
deviates earlier from the correct quantum result.
Comparing the random and discrete sampling results of SM-based mappings in
Figs.~\ref{fig:3site12_RS}a and \ref{fig:3site12_RS}b, we also notice that the
discrete mapping based on $N=8$ trajectories leads to a better description of
the oscillations of the populations than the random-sampling results, since
fewer trajectories hinder dissipation (through averaging) in the classical phase-space.

\begin{figure} \includegraphics[width=0.95\columnwidth]{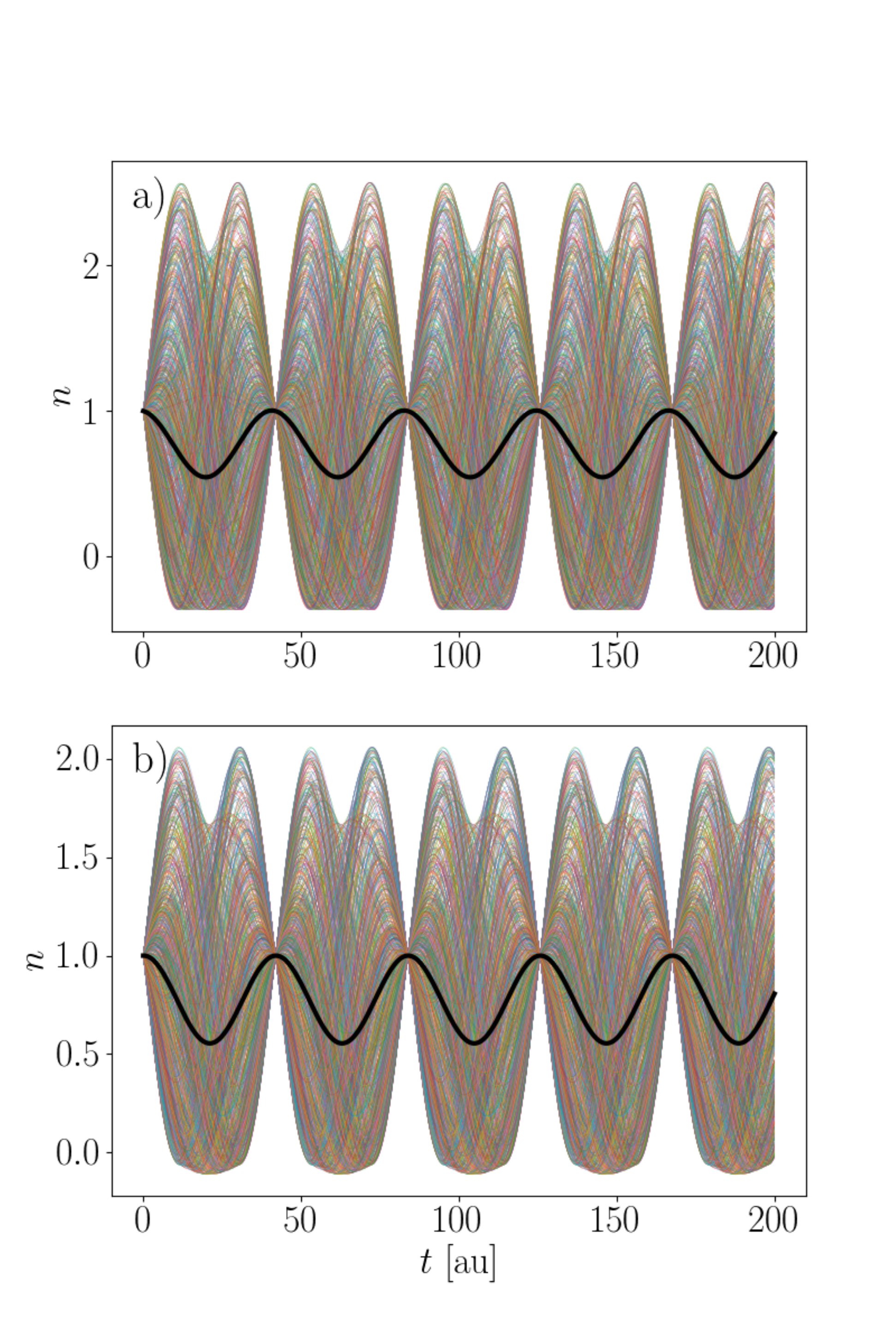}
    \caption{\label{fig:3site12_n0_dist}
    Time-dependent population of site 1, $n_{1\alpha}$, for each single trajectory
    in the 3-sites cyclic tight-binding system for
    a) the MM model and b) the LMM model.
    Both sets of trajectories yield the same exact expectation value of the population for
    this non-interacting Hamiltonian. The expectation
    value $\langle n_{1\alpha}\rangle$ over the trajectories is shown as a thick black line.}
\end{figure}

Finally, we want to examine the population dynamics in the 3-site system from
the perspective of the \emph{individual} trajectories in the random-sampling
case. The population of site 1 for each of the 1000
trajectories
can clearly grow above $n_1=1$ both for the LMM
and MM mappings as seen in Fig.~\ref{fig:3site12_n0_dist}. Nonetheless these
trajectories reproduce in average the exact population dynamics.
The common wisdom is that fermionic mappings need to limit by construction the
maximum value of the classical action variables at the trajectory level
because there can be no more than one fermion per spin-orbital.
We see, however, that imposing this restriction at the level of individual trajectories
is not required for many-body systems without interactions, as only the average value from all trajectories
can be given a physical meaning. Even the fermion-tailored LMM can
reach populations larger than 1 (cf. Fig.~\ref{fig:3site12_n0_dist}b) and still yield the exact
averaged population dynamics.

\begin{figure*}[t!]
	\includegraphics[width=0.95\textwidth]{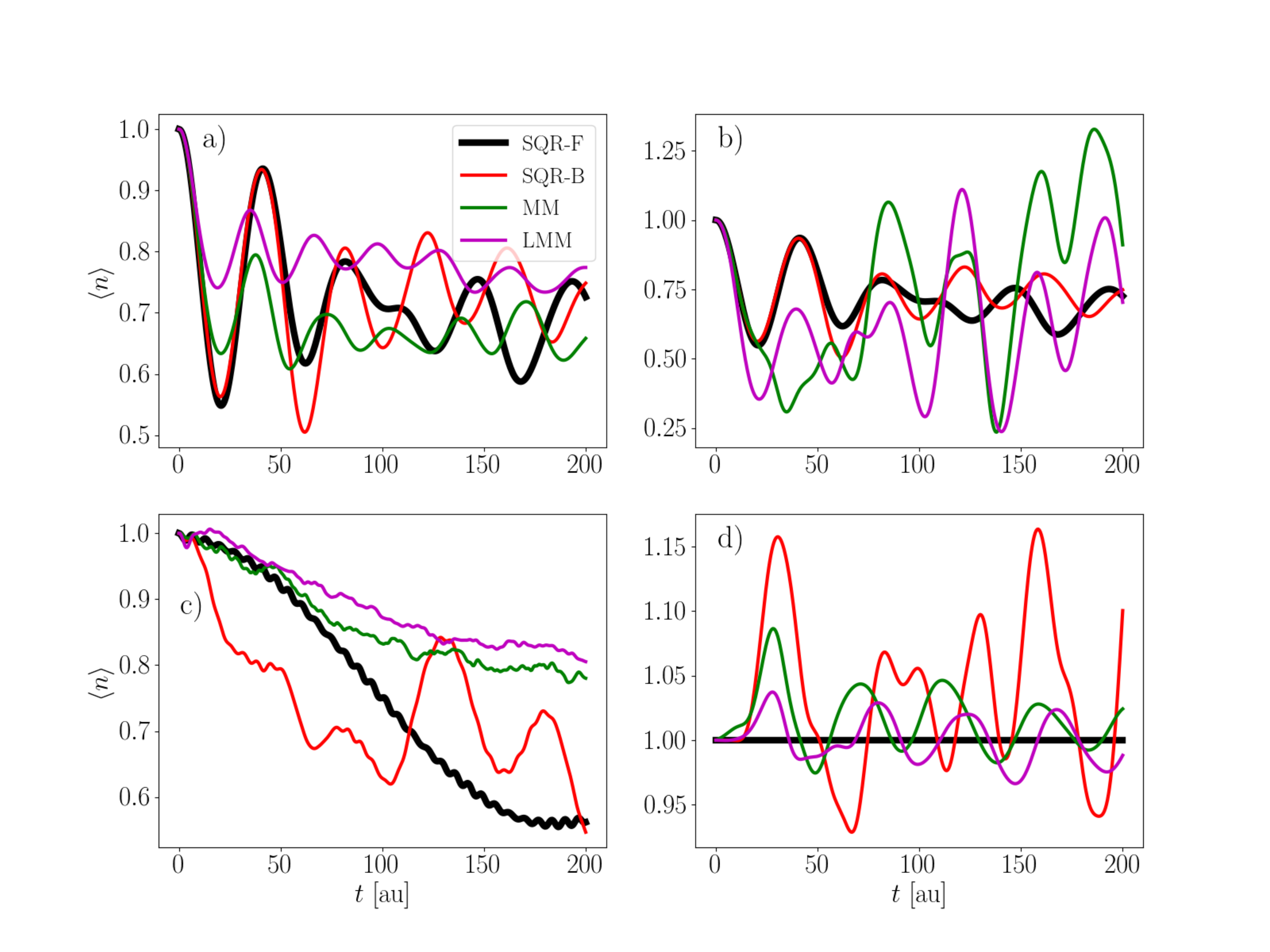}
    \caption{\label{fig:3site12_hub}
    Population dynamics of $n_{1\alpha}$ for the 3-site cyclic
    Hubbard model with $T=-0.05$ and $U=0.05$, except c) where $U=0.5$.
    a) uniform sampling;
    b) discrete sampling;
    c) strong interaction;
    d) all spin-orbitals initially populated.}
\end{figure*}


\subsection{Interacting systems} \label{sec:inter}

In the previous section, we have discussed why the MM mapping
reproduces the exact electron dynamics in non-interaction systems,
whenever the initial phase space distribution matches the initial one-body
density of the system.
Next, we benchmark the performance of the MM mapping in interacting systems against
other mappings, and against exact quantum dynamical results. 
We consider first the Hubbard Hamiltonian,
which consists of a tight-binding Hamiltonian plus an on-site repulsion
$U \hat{c}^\dagger_{i,\alpha}
\hat{c}_{i,\beta} \hat{c}^\dagger_{i,\alpha} \hat{c}_{i,\beta}$.
The classical mapping for the Hubbard interaction term takes the form
$U n_{i\alpha}n_{i\beta}$, where
$n_{i\sigma}$ are functions of the classical phase-space variables.
We first consider a cyclic chain with on-site repulsion terms followed by the
simulation of different impurity Hamiltonians. 
%
%
Afterwards, we compare the mappings in their ability to describe excitonic energy
transfer between model chromophores.


\subsubsection{3-site cyclic system}

\begin{figure*} 
	\includegraphics[width=0.95\textwidth]{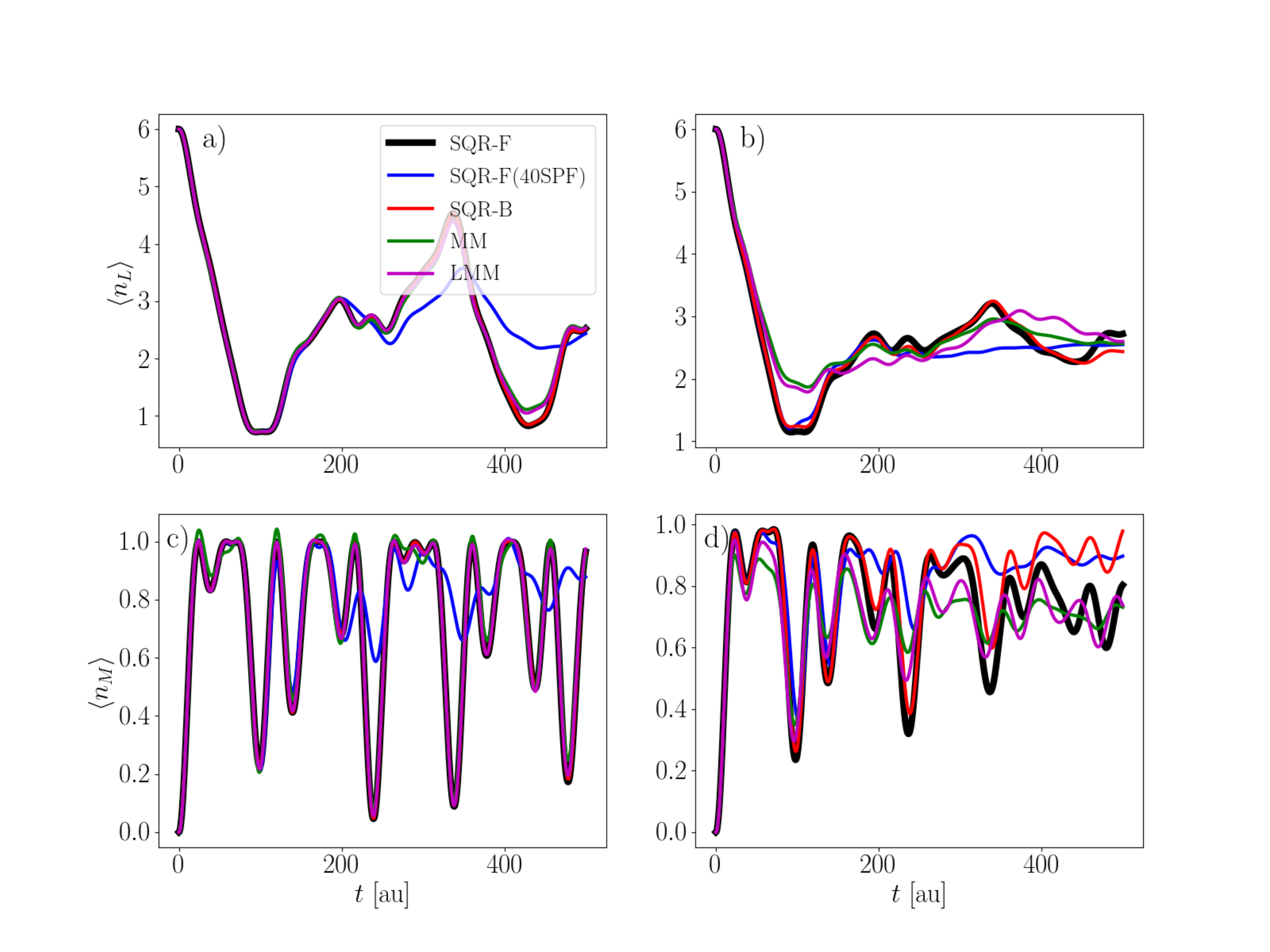}
	\caption{\label{fig:7site} In a linear chain system($N=3$), a) \& b) $n_L$ and c) \& d) $n_M$ are described. In the system, a) \& c) $(T,U)=(-0.05,0.005)$ while b) \& d) $(T,U)=(-0.05,0.05)$. For all classical mappings, trajectories are generated by the uniform sampling.}
\end{figure*}

\begin{figure*} 
	\includegraphics[width=0.95\textwidth]{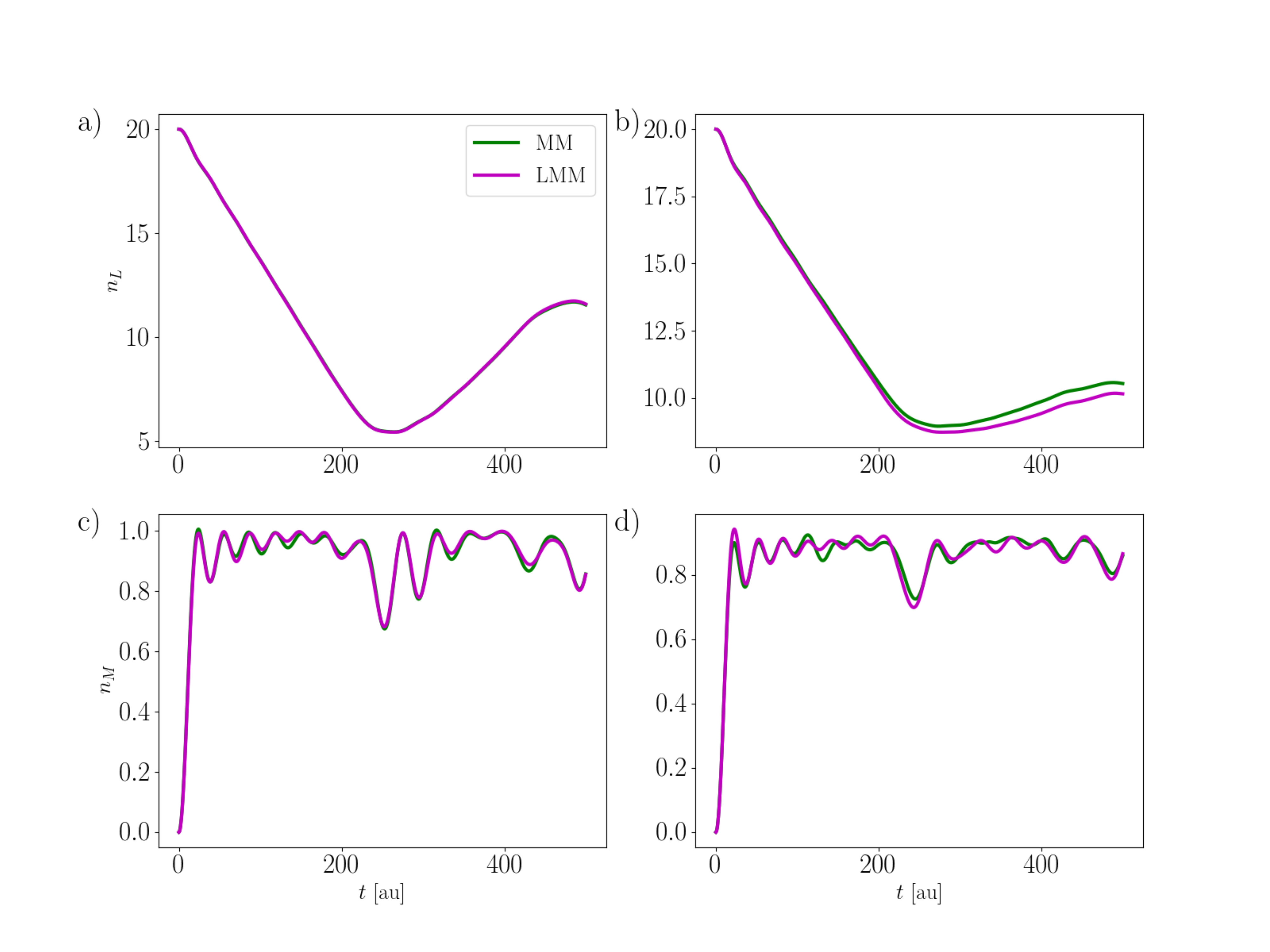}
	
	\caption{\label{fig:21site} In a linear chain system($N=10$), a) \& b) $n_L$ and c) \& d) $n_M$ are described. In the system, a) \& c) $(T,U)=(-0.05,0.005)$ while b) \& d) $(T,U)=(-0.05,0.05)$. For all classical mappings, trajectories are generated by uniform sampling.}
\end{figure*}

Figure~\ref{fig:3site12_hub}a shows the time evolution of the $n_{1\alpha}$
population in the 3-site cyclic system for the ratio $|U/T| = 1$ between the
transfer integral and the on-site repulsion.  The initial state is the same as for
the bosonic and fermionic exact dynamics (cf. Fig~\ref{fig:topo_site}b) and
their population dynamics remains practically identical for more than one period
of oscillation. Naturally, fermionic and bosonic
population dynamics diverge as time progresses and follow a
completely different time-evolution already after about two periods of oscillation.
Both the MM and LMM mappings capture the period of the first oscillation, and
the MM mapping follows more closely the trace of both exact fermions and bosons.
Incidentally, the MM mapping follows more closely the fermionic than the bosonic
population, whereas the LMM mapping is closer to the bosonic than to the
fermionic trace after a few oscillations.
These results constrained to this particular system and parameters and they should not be
overinterpreted. However, they provide two important indications that we are going
to explore with further numerical examples:
i) The MM mapping, which one could consider as the genuine mapping for bosons,
does not necessarily reproduce the bosonic dynamics better than the fermionic
one.
ii) The LMM, which is devised to reproduce certain aspects of fermionic systems,
can indeed in some instances yield results closer to the bosonic
time evolution and does not
necessarily reproduce the fermionic dynamics more accurately than the simpler MM
prescription.

We have seen how, in the absence of interactions, \emph{any} set of
trajectories that exactly maps the initial one-body density of the quantum system
yields the exact evolution of the
populations and one-body correlations.
For interacting systems this is not the case anymore. Different initial
ensembles encode different two-body densities, and these evolve differently under
interactions.
The population dynamics in Fig.~\ref{fig:3site12_hub}b corresponds to the same
parameters as in Fig.~\ref{fig:3site12_hub}a but now the initial density one-body
density is mapped with only $N=8$ trajectories, instead of a uniform sampling.
Both MM and LMM mappings are found to perform worse in comparison with the quantum
mechanical results than the uniform sampling results.
However, compared to the uniform sampling, the ensemble dynamics of the
discrete sampling retains oscillations for a longer time because a smaller
number of phase-space points are being averaged.
This indicates that a possible strategy to
improve on these results might involve mapping both the one- and two-body
densities of the quantum system to the smallest possible number of
trajectories, but
we are not pursuing this strategy here and instead use the uniform sampling
strategy in the following.

In the strong repulsion limit, $U>>T$
the fermionic and bosonic exact quantum dynamics in Fig~\ref{fig:3site12_hub}c quickly diverge.
It is surprising that both MM and LMM mappings still reproduce the
fermionic dynamics rather well at early times, and we
have not been able to develop an intuitive explanation for this fact.
An extreme example consists of a system with all spin-orbitals occupied. Under
fermionic statistics, such a system is blocked: all its orbital
populations remain equal to 1 at all times.
Its bosonic counterpart, though, does not experience
this Fermi blockade and its populations can present fluctuations.
These
fluctuations vanish for a non-interacting system,
or in the case that the transfer integrals and on-site repulsions are fully symmetric.
Figure~\ref{fig:3site12_hub}d shows the populations of the fully-occupied Hubbard
system with $|U/T|=1$ (with $U=0$ for site 1 only). Whereas the quantum fermionic
populations remain constant, the bosonic populations fluctuate and so do the
populations calculated with both mappings to within a similar range. Intriguing are
the fluctuations of the LMM mapping, which is devised to reproduce the behaviour of
fermions. This indicates again that the LMM mapping is not superior to the simpler
MM prescription at reproducing the dynamics of a fermionic system, and they perform
similarly in this connection.

\subsubsection{Impurity Hamiltonian}

Impurity models describe electron transport processes and have been approached
by classical mappings in recent works~\cite{Levy2019}.
Here we consider two tight-binding chains (left and right) of $N$ conduction orbitals
coupled to a central single-impurity orbital with a local interaction term $U$ (cf. Fig.~\ref{fig:topo_site}c)
In total each model consists of $2N+1$ sites and we consider chains with $N=3,10$.

Simulations in the $N=3$ system (7 sites) can still be easily performed quantum mechanically and
comparisons with the classical mappings are shown in Fig~\ref{fig:7site}. Simulations with
$N=10$ are only approached using the classical mappings.
We consider weak and stronger interaction regimes, $|U/T|=0.1$ and $|U/T|=1$, respectively, and set
all electrons to be on the left conduction chain of the impurity model at $t=0$. These are quite
extreme initial conditions in terms of chemical potential if one compares with simulations aimed
at reproducing steady-state conditions of the impurity model~\cite{Levy2019}. We emphasize that
our goal is to benchmark and characterize the performance of the mappings against full quantum results,
rather that reproduce specific experimental conditions.

The population (number of electrons) of the left conduction chain and of the central impurity site
for the $N=3$ are shown for the two coupling strengths in Fig~\ref{fig:7site}.
The quantum fermionic and bosonic dynamics are similar for both coupling strengths when starting from
a fermionic initial state.
%
%
The exact fermionic calculations require 64 single particle functions (SPF) to span the corresponding
sub-Fock spaces of the left and right
conduction chains and 4 SPFs for the middle site~\cite{sas2020:154110}. The truncated fermionic calculation
uses a basis of 40 SPFs for the left and right sites and 4 SPFs for the middles site.

In the case of weak interaction, both the quantum bosonic and fermionic dynamics (red and black traces) remain
very similar and both the MM and LMM mappings reproduce the population dynamics extremely well, as seen
in Fig~\ref{fig:7site}a and Fig~\ref{fig:7site}c.
The worst result corresponds to the non-exact quantum mechanical calculation with 40 SPFs (blue trace).
The reason for this is easily understood. In calculations based on a second-quantization
representation~\cite{wan09:024114, sas2020:154110},
the correlation between the computational degrees of freedom (i.e. the orbital occupations) depends on the hopping
integrals $T$, not on the on-site electron repulsion $U$. Indeed, the case of stronger electron-electron repulsion
is comparatively better described by the approximate quantum calculation as compared to the exact result. 

Even in the case of stronger coupling, $|U/T|=1$, both classical mappings are able to reproduce the population
of the conduction chains and impurity site remarkably well (Figs.\ref{fig:7site}b and \ref{fig:7site}d).
Interestingly, the population of the impurity never surpasses $\langle n_M\rangle =1$ for either classical mapping
even when the electronic
flux towards the right conduction chain is maximal during the first 100~au of time.
This again strengthens the observation that enforcing a
limitation of the maximal population of the fermionic degrees of freedom via the classical mapping (e.g. using
classical spin DOFs) is unnecessary, and that the MM performs similarly as LMM while using half the number of classical
DOFs.

Finally, we consider an impurity
model with $N=10$ sites in each conduction chain and show its population dynamics,
in Fig~\ref{fig:21site}.
where
we compare the MM and the LMM mappings for both interaction strengths.
As the number of sites increases, the averaged populations of the left, central and right sites becomes more similar
between both mappings. The small fluctuations of the central impurity in Figs.~\ref{fig:21site} are perfectly captured
and the population of the impurity does not grow above 1.

\subsection{Excitonic energy transfer}

In the previous section, we have shown that the MM mapping can be accurate in systems with weak and medium interactions. 
Here, we benchmark the applicability of the MM mapping to
a second-quantization model for inter-molecular energy transfer between electronically
excited molecules mediated by
dipole-dipole interactions, i.e. excitonic energy transfer (EET).
Although EET has been very successfully described by variational full quantum simulations, for example
based on the MCTDH approach~\cite{tam15:107401},
realistic excitonic complexes may consist of many
thousands of degrees of freedom,
in which case mapping-based approaches an might be a useful alternative.

\begin{figure}
	\includegraphics[width=0.95\columnwidth]{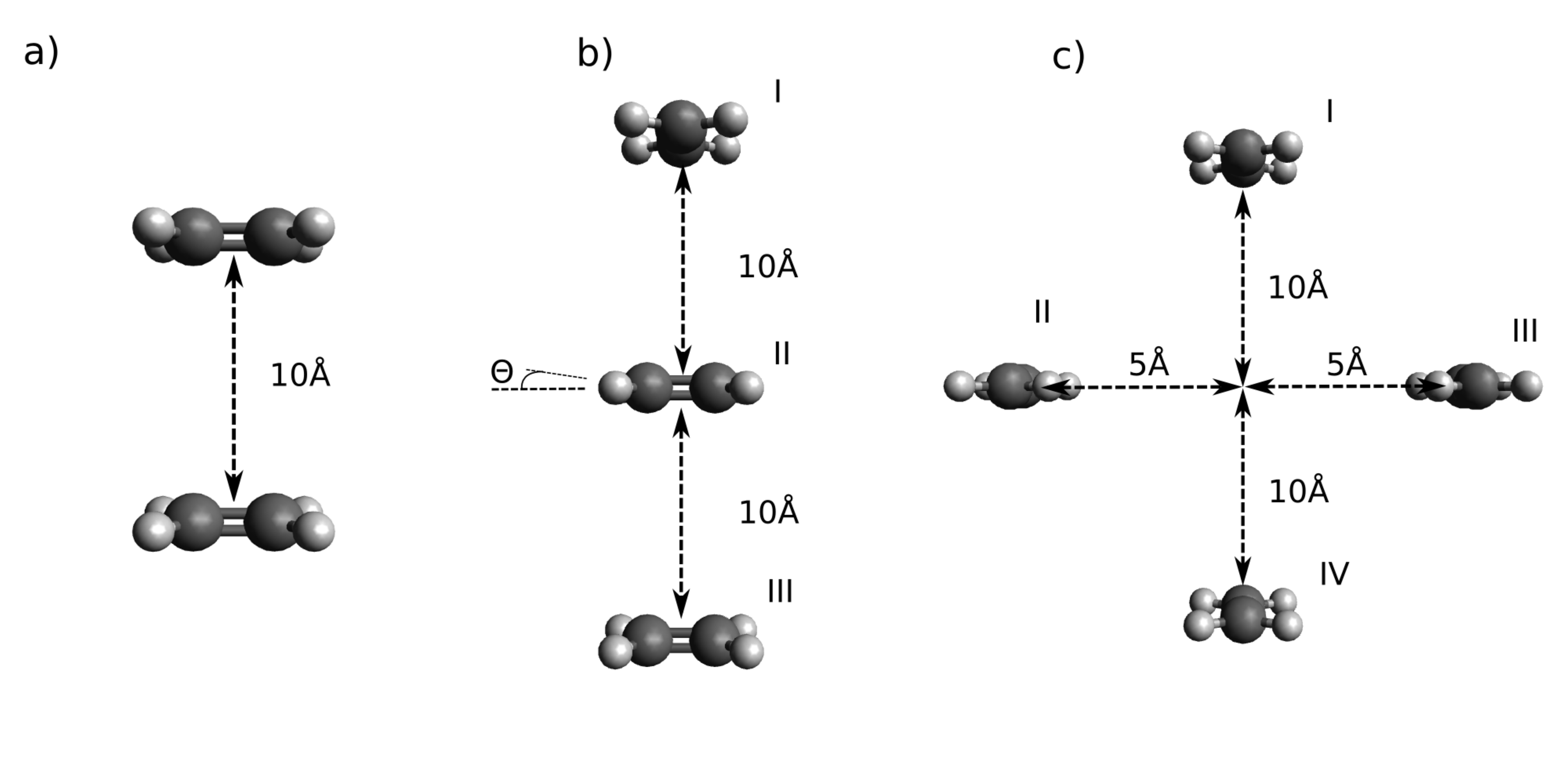}
	\caption{\label{fig:Ethy_st} Ethylene a) dimer, b) trimer, and c) tetramer used in the EET simulations.}
\end{figure}

We benchmark the MM and LMM mappings to EET on ethylene clusters (Fig.\ref{fig:Ethy_st}) in various configurations,
while keeping the nuclei fixed.
Each ethylene molecule is described within an orbital approximation and
the construction of the EET model parameters for each cluster proceeds as follows:
Localized molecular orbitals are obtained for each ethylene molecule through a separate Hartree-Fock calculation using
a minimal atomic basis.
Only the highest occupied molecular orbital (HOMO) and lowest occupied molecular orbital (LUMO) in each subsystem
are further considered to describe the EET. Hence, each model consists of 4$N$
spin-orbitals, where $N$ is the number of ethylene molecules.
The electrons are considered independent within each monomer and interact with the electrons of the
other monomers via the two-electron Coulomb integrals involving
the HOMOs and LUMOs of each monomer pair. Exchange terms are negligible
due to distance between the monomers.
The one- and two-body electronic integrals for each cluster configuration are provided as supporting
information.



\begin{figure}
	\includegraphics[width=0.95\columnwidth]{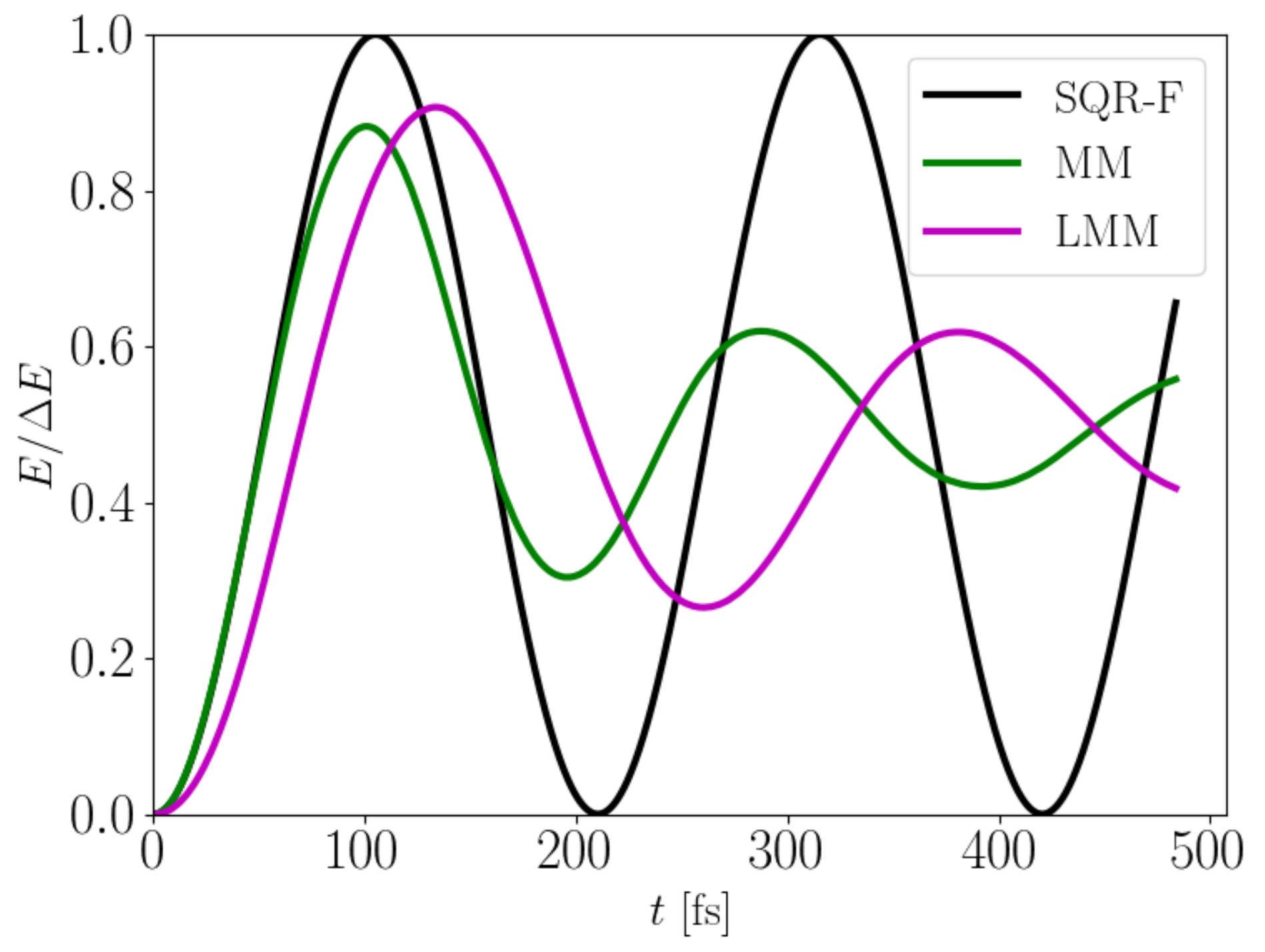}
	\caption{\label{fig:Ethy} Energy (in units of the HOMO-LUMO energy gap $\Delta E$)
	of the acceptor ethylene molecule in the ethylene dimer system.}
\end{figure}

We first consider two ethylene molecules facing each other and separated by 10~{\AA} (Fig.\ref{fig:Ethy_st}a).
The initial state consists of one of the molecules singly excited (HOMO$\to$LUMO) and the other molecule in its
ground electronic state. The two localized excitonic states are resonant, resulting in a simple periodic EET
with a period of about 200~fs. The direct Coulomb repulsion integrals between HOMO and LUMO orbitals in different
monomers have values of roughly 1~eV, which contribute to energy shifts of the orbitals.
The Coulomb integrals of the form
$V_{L_I,H_{II}}^{L_{II},H_I}$ directly drive the exchange of excitation between the monomers and
have a value of roughly 50~cm$^{-1}$.
The EET dynamics is approximately captured by the MM and LMM mappings during the first period
as seen in Figure~(\ref{fig:Ethy}), where the energy of the acceptor molecule is shown in units
of the HOMO-LUMO energy gap.
Both mappings fail to capture the full recursions of the exact dynamics due classical phase
space dissipation of the ensemble of trajectories.
Crucially, however,  MM reproduces the initial rate of energy
transfer while LMM deviates from the exact curve almost from the onset.

%

Next, we consider the situation where the EET proceeds through a bridging molecule (Figure~(\ref{fig:Ethy_st} b)).
The relative orientation
between the C-C axis of the donor/acceptor and bridge molecules determines the strength of the dipole-dipole
coupling, being it equal to zero for an orthogonal configuration. In the studied cluster, the donor and acceptor systems
have a relative orientation of 90 degrees and therefore there is zero energy transfer between them without the
intervention of the bridging molecule. A similar system was considered, e.g., in Ref.~\citenum{Ford2014}.
The bridging system is considered at three angles $\theta=0,30,45$ with respect to the acceptor molecule.
For $\theta=0$, no EET can occur because the transition dipole moments between the donor and the
acceptor, as well as the mediator and the acceptor, are zero.
For $\theta=30$~deg., partial EET to the acceptor becomes possible. This partial EET is captured by both the
MM and LMM mappings. In both cases, the classical models yield a smaller amount of energy transfer by 10 to
20\% with respect to full quantum results after the first half period.
As seen in Figs.~(\label{fig:Ethy_tri}a, \label{fig:Ethy_tri}b),
the time-scale of the first period is well captured by the MM mapping, whereas LMM again yields
a slower EET dynamics than the exact result.
The general trends are similar for $\theta=45$~degrees. The EET to the acceptor molecule is almost
complete after the first half period. Both classical models yield a smaller total energy transfer by
about 20\% and LMM again yields a slower EET dynamics than the exact result.

\begin{figure}
	\includegraphics[width=0.95\columnwidth]{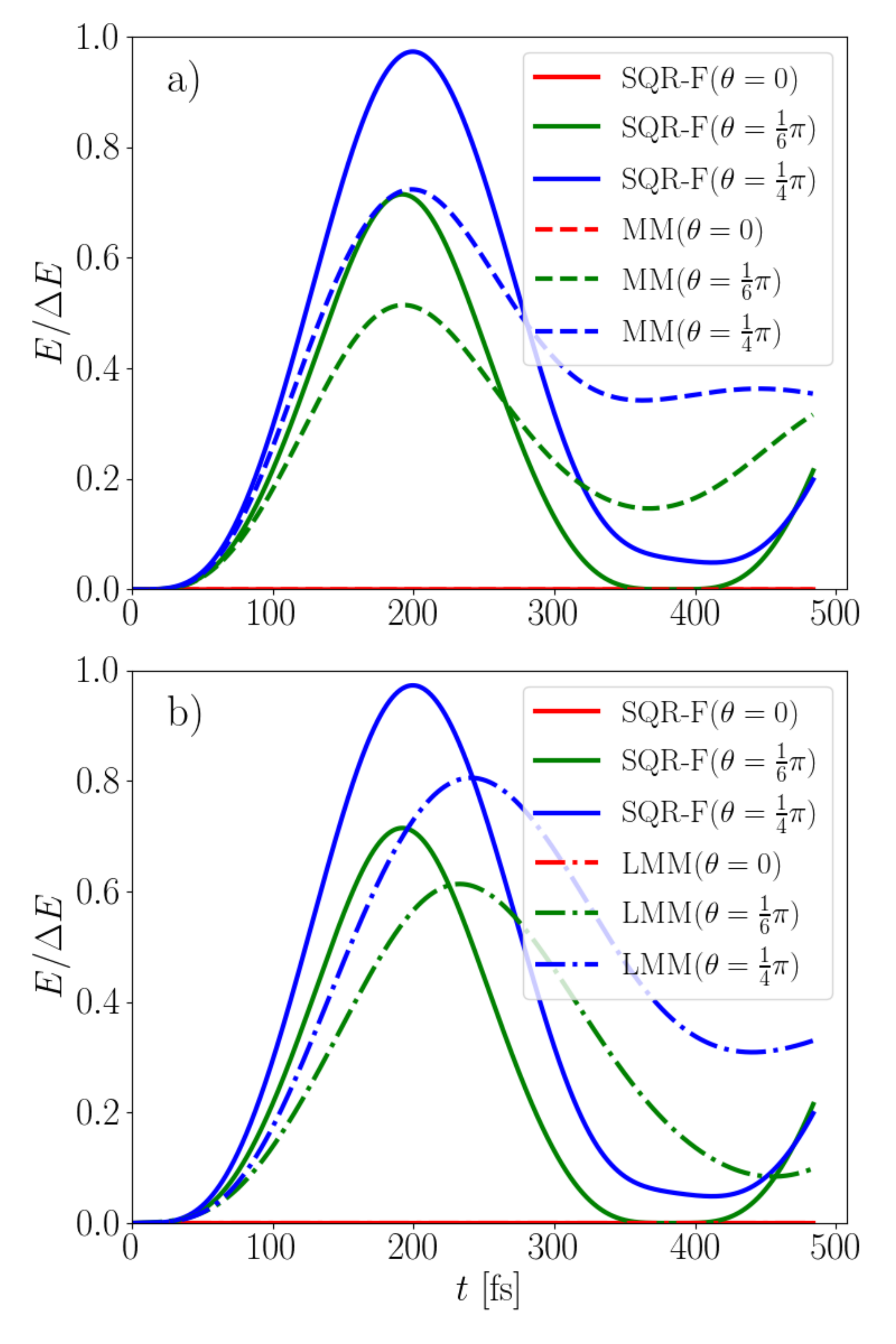}
	
	\caption{\label{fig:Ethy_tri}  Energy of the acceptor monomer in the ethylene trimer system (normalized to the HOMO-LUMO energy gap). The donor is initially excited to the first singlet excited state.
	a) MM mapping and b) LMM mapping compared to the exact quantum mechanical result for different
	relative orientations of central
	monomer.}
\end{figure}

\begin{figure}
	\includegraphics[width=0.95\columnwidth]{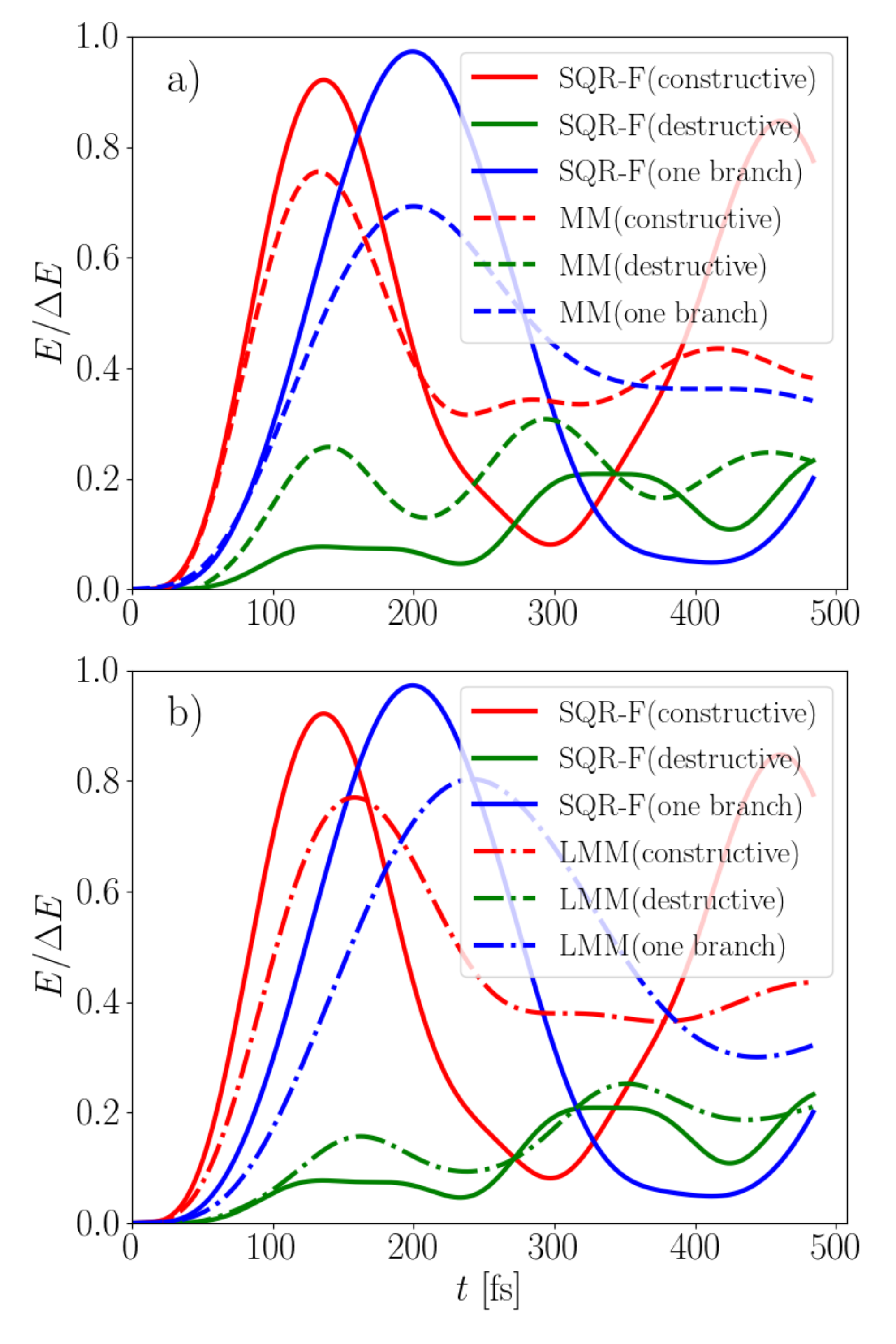}
	
	\caption{\label{fig:Ethy_tet} Energy of the acceptor monomer in the ethylene tetramer system (normalized to the HOMO-LUMO energy gap). The donor is initially excited to the first singlet excited state.
	a) MM mapping and b) LMM mapping compared to the exact quantum mechanical result.}
\end{figure}

Finally, we consider a cluster with four monomers, two acting as donor (I) and
acceptor (IV) systems, while the two other monomers (II, III) act as a symmetrical
bridge between the former two.
In all cases, monomer I is initially excited at $t=0$ and
the energy of monomer IV as a function of time is 
shown in Figure~(\ref{fig:Ethy_tet}). 
The EET is faster when the two bridging pathways constructively
interfere. These kinds of coherent EET dynamics involving various 
pathways are known to operate in models of light-harvesting
complexes~\cite{tom20:2348}. The onset of the EET process is well captured
by the MM mapping, while the EET described by LMM is slower
(cf. slopes during first 100~fs in Figs.~(\ref{fig:Ethy_tet}a) and (\ref{fig:Ethy_tet}b)).
After the first
period, about 120~fs, both mappings deviate from the exact result and do
not capture the almost complete back-transfer to monomer I after about 270~fs.
When one of the pathways is completely suppressed, the EET process also
reaches almost 100\% yield, although it requires now a longer time, roughly 220~fs, until monomer IV is
fully excited. Similarly as before, the initial EET dynamics and the duration of
the first half period are captured by the MM mapping.
For all the considered ethylene clusters, the MM mapping properly describes
the EET rate while the LMM underestimate the EET rate in the first half
period.

Next, we artificially change the sign of the coupling matrix element between
monomers II and IV. This results in a destructive interference and
the EET process is strongly suppressed, reaching only about 10 to 20\% of
the EET yield compared to the constructive case.
Interestingly, the classical mappings capture this destructive interference effect
and describe a strongly suppressed EET process. In this example, LMM reproduces the
exact quantum result slightly better than MM at short times,
and both mappings describe the EET dynamics approximately to within 10 to 20\% of the
HOMO-LUMO energy gap for the whole duration of the simulation, about 500~fs.

\section{Conclusions}



We have bench-marked and
analyzed the application of various classical mappings (MM, SM, SM+JWT, LMM) to describe
the dynamics of fermions (electrons) in Fock space. Our goal has been to establish the
range of applicability and possible drawbacks of the MM mapping and to compare it in detail
with the LMM mapping, which includes design principles tailored to the classical description
of fermionic dynamics in Fock space.

For non-interacting particles, the MM mapping yields an exact description of
the one-body density of the quantum system, provided the dynamics starts
from a state with fermionic occupation. This occurs because
non-interacting fermionic and bosonic systems starting from identical one-body
densities have the same one-body dynamics (cf. Eq.~(\ref{eq:pops})).

Mapping the quantum one-body density to the classical phase space always
requires an ensemble of trajectories, the only exception being the case in
which only one electron is present. The phase-space mapping can be achieved
either by randomly sampling the classical angle coordinates while
selecting action-coordinate values that fulfill the initial quantum populations,
or by picking a discrete set of points that yield the quantum one-body density when
averaged under Eq.~(\ref{eq:nij_pq_int}).
Both schemes are equivalent in the non-interacting case.

The SM mapping, with or without a Jordan-Wigner transformation,
fails to describe the dynamics of non-interacting systems. This fact was already
known and is due to the non-linear nature of Hamilton's equations originating
from these models~\cite{Li2012,Li2013,Li2014}.
Importantly, combining the Jordan-Wigner
transformation with the SM, which has been shown to be
equivalent to the original MW mapping~\cite{Miller1986},
does not result in any improvement.

We further considered the LMM mapping for fermions in
comparison with the simpler MM mapping. LMM uses two
vectorial coordinates per fermionic degree of freedom to restrict its
maximal occupation and accounts for the sign-rules of fermionic
matrix elements through the form of the
interacting terms of the classical analog.
LMM is in spirit similar to the original
MW mapping.
It can be obtained by mapping each fermionic DOF to two harmonic oscillators
plus including the classical functions of the sign-change operators
of a Jordan-Wigner transformation.
LMM yields linear equations of motion for the non-interacting case and hence
the exact one-body dynamics, but it uses twice as many degrees of freedom
as the simpler MM mapping.
These two mappings were applied to Hubbard-like cyclic models, to
linear-chain impurity models and to models of excitonic energy transfer.
Cyclic models provide important insights in the role of anti-symmetry
in classical mappings for fermions. This is because the one-body
part of the Hamiltonian contains sign-change operators after the JWT, whereas
they cancel out in the corresponding linear models.
We found that inclusion of these non-linear terms in the MM (and SM) mapping does more harm
than good and worsens the description in all considered examples.
As mentioned, linear tight-binding and Hubbard Hamiltonians do not feature
sign-change operators after the JWT. Hence, the only actual difference between
MM and LMM mappings in these examples is the limitation of the maximal orbital
occupation in the LMM case. However, this enforced limitation of the LMM
mappping plays no particular role in the considered impurity models. The orbital
occupation in the central impurity site is very similar in both mappings and
always below the maximal occupation of 1 electron per spin-orbital.

Finally, we considered the process of excitonic energy transfer in model clusters of ethylene.
In the models, monomers interact via the two-body integrals calculated from localized
Hartree-Fock orbitals in each monomer. We found that the MM mapping correctly captures the
initial EET rate (roughly the first half period) and later on deviates from the exact dynamics. LMM
leads to a qualitatively similar dynamics compared to MM, although the initial EET rate is often
slower than the exact bench-mark result. Both classical mappings correctly capture constructive and
destructive interferences resulting from equivalent energy transfer pathways between the donor and
acceptor monomers.

Our results indicate that the original MM mapping may be a valid classical
analog alternative for the description of fermions in Fock space, and that it
does not perform worse than mappings designed specifically for fermions. Our
results also cast a question mark on whether the Jordan-Wigner transformation
(or an equivalent formulation) is a useful addition to classical mapping
strategies for fermions.  Our conclusion is that it should be avoided. A
limitation by construction of the maximal fermionic occupation does not seem to
be necessary either, at least within the bench-marked examples. In fact,
in most of the considered examples,
the MM mapping outperforms the more complicated LMM.
Future work shall consider sampling strategies of the initial phase-space
distribution designed for fermions, i.e. explicitly solving
Eqs.~(\ref{eq:nij_actang}) for a discrete set of phase-space points, as well as
the inclusion of nuclear displacements.

\section*{Supplementary material}
See the supplementary material for the 1-eletron and 2-electron integrals in ethylene clusters.

\begin{acknowledgments}
    J.S. acknowledges the International Max-Planck Research School for Quantum Dynamics (IMPRS-QD)
    for financial support.
    The authors are grateful to Prof. Dr. H.-D. Meyer for his valuable suggestions during the
    preparation of the manuscript and for his critical advice.
\end{acknowledgments}

\section*{Data availability}
The data that support the findings of this study are available from the corresponding author 
upon reasonable request.

\section*{Appendix}

\subsection{Classical mappings for fermions}

A major obstacle in the classical description of fermions is the anti-commutation relation of the
fermionic creation and annihilation operators~(\ref{eq:commutation}).
The anti-commutation relation has two main consequences: it limits the maximum of occupation of the fermionic
degrees of freedom in the occupation-number representation and it introduces a $\pm 1$ phase to the
corresponding matrix elements depending on the occupation pattern or the orbitals.
Classical analogs for fermions attempt to include these properties using different strategies.
The orbital occupation can be limited through the use of spin or angular momentum degrees of freedom,
whereas the phase of the matrix elements can be included via a Jordan-Wigner transformation from
fermionic to spin degrees of freedom followed by the introduction of the classical analog.
The form of these mappings for non-interacting systems is introduced in the following
for comparison with the MM mapping, and the reader is referred to the
original references where the derivations and interacting terms are discussed in detail.

\subsubsection{SM mapping}

The SM mapping has already been utilized as a classical analog for electronic degrees of freedom
(DOF) for a finite set of electronic states~\cite{mey79:3214,cot15:12138,Meyer1979_2,Meyer1980}
and for interacting fermions~\cite{Swenson2011}.
One proceeds by first mapping the fermionic creation-annihilation operators to the angular momentum operators for
spin $1/2$ degrees of freedom
\begin{align}
\begin{split}
    \label{eq:c-to-sxyz}
    \hat{c}_i & \mapsto \hat{\sigma}_{i,x} + i\hat{\sigma}_{i,y}\\
    \hat{c}_i^{\dagger} & \mapsto \hat{\sigma}_{i,x} - i\hat{\sigma}_{i,y},
\end{split}
\end{align}
 where $\hat{\sigma}_{i,z} = 1/2$ ($\hat{\sigma}_{i,z} = -1/2$) corresponds to the occupied (empty) $i-th$ one-particle state.
The classical Hamiltonian function (for non-interacting particles) without JWT reads  
\begin{align}
\begin{split}
    \label{eq:h1cl_sm}
    {H}^{(1)}_{\text{SM}}({\sigma}_{x}, {\sigma}_{y}, {\sigma}_{z}) &= 
        \sum_{i}h_{ii}\left({\sigma}_{z,i}+\frac{1}{2}\right)\\
        &+2\sum_{i>j}h_{ij}\left({\sigma}_{x,i} {\sigma}_{x,j}+ {\sigma}_{y,i} {\sigma}_{y,j}\right).
\end{split}
\end{align}
with classical equations of motion~\cite{cot15:12138}
\begin{align}
    \label{eq:hamilton_sm}
    \frac{d\overrightarrow{\sigma_{i}}}{dt} &= \frac{\partial H(\overrightarrow{\sigma_{i}})}{\partial\overrightarrow{\sigma_{i}}} \times \overrightarrow{\sigma_{i}}\\
    \overrightarrow{\sigma_{i}} &= (\sigma_{x,i}, \sigma_{y,i}, \sigma_{z,i}) .
\end{align}

On the other hand, the Jordan-Wigner transformation (JWT) exactly maps a fermionic Hamiltonian to a spin-chain Hamiltonian
that yields matrix elements with the correct fermionic phase~\cite{Jordan1993}.
Thus, it seems reasonable to combine the SM with the JWT, which introduces phase operators
$\hat{S}_k = \exp(i\pi\hat{n}_k)$,
and then introduce the corresponding classical functions.
These can take the form
\begin{equation}\label{eq:sg_oper}
S_k = (1-2n_{k}),
\end{equation}
where the occupation $n_j(\vec{x})$ depends on the explicit classical variables $\vec{x}$ used in the
corresponding mapping.
The SM+JWT classical analog Hamiltonian for non-interacting particles then takes the form
\begin{align}
\begin{split}
    \label{eq:h1cl_sm_sc}
    {H}^{(1)}_{\text{SM+JWT}}({S}_{x}, {S}_{y}, {S}_{z}) &= 
        \sum_{i}h_{ii}\left({S}_{z,i}+\frac{1}{2}\right)\\
        &+2\sum_{i>j}\prod_{k=j+1}^{i-1} h_{ij}  (1-2n_{k})\left({S}_{x,i} S_{x,j} + {S}_{y,i} {S}_{y,j}\right).
\end{split}
\end{align}
Comparing Eq.~(\ref{eq:h1cl_sm_sc}) with the relation 2.10 in Ref.~\citenum{Miller1986} shows
that this procedure is equivalent to the original Miller-White mapping.

\subsubsection{LMM mapping}

The Li-Miller mapping (LMM)
uses the concept of quaternion operators to capture the properties of second-quantized fermionic operators and
to construct a classical mapping for them~\cite{Li2012,Li2013,Li2014}. This results in a mapping where
each fermionic DOF is mapped to two classical DOFs (two position-momentum pairs).
The LMM has been introduced in the context of semi-classical initial-value representation calculations mechanics~\cite{Li2012,Li2013,Li2014}, whereas here we use it in a fully linearized context.
Recent applications of a classical analog to quantum transport of fermions have modified the LMM
mapping in a way that makes it possible describe a few extra types of fermionic operator products, and
termed it complete quasicalssical map (CQM)~\cite{Levy2019}. In our applications, the LMM and CQM maps
are equivalent.
The corresponding Hamiltonian reads
\begin{align}
    \label{eq:h1cl_cqm}
    {H}^{(1)}_{\text{LMM}}(\mathbf{p}_{x}, \mathbf{q}_{x}, \mathbf{p}_{y}, \mathbf{q}_{y}) =
    \frac{1}{2} \left( \mathbf{q}_{x}^+ \, \mathbf{h}\, \mathbf{p}_{y} - 
                       \mathbf{q}_{y}^+ \, \mathbf{h}\, \mathbf{p}_{x} \right)
                    + \gamma \mathrm{Tr}[\mathbf{h}].
\end{align}
The corresponding Hamilton equations of motion for a non-interacting Hamiltonian read
\begin{align}
\begin{split}
    \label{eq:CQM_eom}
    \overset{\bullet}{\mathbf{q}_{x}} &= -\mathbf{h}\, \mathbf{q}_{y};\quad
        \overset{\bullet}{\mathbf{p}_{x}} = -\mathbf{h}\, \mathbf{p}_{y}\\
    \overset{\bullet}{\mathbf{q}_{y}} &= \mathbf{h}\, \mathbf{q}_{x};\quad
        \overset{\bullet}{\mathbf{p}_{y}} = \mathbf{h}\, \mathbf{p}_{x}.
\end{split}
\end{align}
%

%
%
%


\begin{thebibliography}{51}%
\makeatletter
\providecommand \@ifxundefined [1]{%
 \@ifx{#1\undefined}
}%
\providecommand \@ifnum [1]{%
 \ifnum #1\expandafter \@firstoftwo
 \else \expandafter \@secondoftwo
 \fi
}%
\providecommand \@ifx [1]{%
 \ifx #1\expandafter \@firstoftwo
 \else \expandafter \@secondoftwo
 \fi
}%
\providecommand \natexlab [1]{#1}%
\providecommand \enquote  [1]{``#1''}%
\providecommand \bibnamefont  [1]{#1}%
\providecommand \bibfnamefont [1]{#1}%
\providecommand \citenamefont [1]{#1}%
\providecommand \href@noop [0]{\@secondoftwo}%
\providecommand \href [0]{\begingroup \@sanitize@url \@href}%
\providecommand \@href[1]{\@@startlink{#1}\@@href}%
\providecommand \@@href[1]{\endgroup#1\@@endlink}%
\providecommand \@sanitize@url [0]{\catcode `\\12\catcode `\$12\catcode
  `\&12\catcode `\#12\catcode `\^12\catcode `\_12\catcode `\%12\relax}%
\providecommand \@@startlink[1]{}%
\providecommand \@@endlink[0]{}%
\providecommand \url  [0]{\begingroup\@sanitize@url \@url }%
\providecommand \@url [1]{\endgroup\@href {#1}{\urlprefix }}%
\providecommand \urlprefix  [0]{URL }%
\providecommand \Eprint [0]{\href }%
\providecommand \doibase [0]{http://dx.doi.org/}%
\providecommand \selectlanguage [0]{\@gobble}%
\providecommand \bibinfo  [0]{\@secondoftwo}%
\providecommand \bibfield  [0]{\@secondoftwo}%
\providecommand \translation [1]{[#1]}%
\providecommand \BibitemOpen [0]{}%
\providecommand \bibitemStop [0]{}%
\providecommand \bibitemNoStop [0]{.\EOS\space}%
\providecommand \EOS [0]{\spacefactor3000\relax}%
\providecommand \BibitemShut  [1]{\csname bibitem#1\endcsname}%
\let\auto@bib@innerbib\@empty
\bibitem [{\citenamefont {Miller}\ and\ \citenamefont
  {White}(1986)}]{Miller1986}%
  \BibitemOpen
  \bibfield  {author} {\bibinfo {author} {\bibfnamefont {W.~H.}\ \bibnamefont
  {Miller}}\ and\ \bibinfo {author} {\bibfnamefont {K.~A.}\ \bibnamefont
  {White}},\ }\href@noop {} {\bibfield  {journal} {\bibinfo  {journal} {J.
  Chem. Phys.}\ }\textbf {\bibinfo {volume} {84}},\ \bibinfo {pages} {5059}
  (\bibinfo {year} {1986})}\BibitemShut {NoStop}%
\bibitem [{\citenamefont {Meyer}\ and\ \citenamefont
  {Miller}(1979{\natexlab{a}})}]{mey79:3214}%
  \BibitemOpen
  \bibfield  {author} {\bibinfo {author} {\bibfnamefont {H.~D.}\ \bibnamefont
  {Meyer}}\ and\ \bibinfo {author} {\bibfnamefont {W.~H.}\ \bibnamefont
  {Miller}},\ }\href {\doibase 10.1063/1.437910} {\bibfield  {journal}
  {\bibinfo  {journal} {J. Chem. Phys.}\ }\textbf {\bibinfo {volume} {70}},\
  \bibinfo {pages} {3214} (\bibinfo {year} {1979}{\natexlab{a}})}\BibitemShut
  {NoStop}%
\bibitem [{\citenamefont {Meyer}\ and\ \citenamefont
  {Miller}(1979{\natexlab{b}})}]{Meyer1979_2}%
  \BibitemOpen
  \bibfield  {author} {\bibinfo {author} {\bibfnamefont {H.}~\bibnamefont
  {Meyer}}\ and\ \bibinfo {author} {\bibfnamefont {W.~H.}\ \bibnamefont
  {Miller}},\ }\href {\doibase 10.1063/1.438598} {\bibfield  {journal}
  {\bibinfo  {journal} {J. Chem. Phys.}\ }\textbf {\bibinfo {volume} {71}},\
  \bibinfo {pages} {2156} (\bibinfo {year} {1979}{\natexlab{b}})},\ \Eprint
  {http://arxiv.org/abs/https://aip.scitation.org/doi/pdf/10.1063/1.438598}
  {https://aip.scitation.org/doi/pdf/10.1063/1.438598} \BibitemShut {NoStop}%
\bibitem [{\citenamefont {Meyer}\ and\ \citenamefont
  {Miller}(1980{\natexlab{a}})}]{mey80:2272}%
  \BibitemOpen
  \bibfield  {author} {\bibinfo {author} {\bibfnamefont {H.-D.}\ \bibnamefont
  {Meyer}}\ and\ \bibinfo {author} {\bibfnamefont {W.~H.}\ \bibnamefont
  {Miller}},\ }\href {\doibase 10.1063/1.439462} {\bibfield  {journal}
  {\bibinfo  {journal} {J. Chem. Phys.}\ }\textbf {\bibinfo {volume} {72}},\
  \bibinfo {pages} {2272} (\bibinfo {year} {1980}{\natexlab{a}})}\BibitemShut
  {NoStop}%
\bibitem [{\citenamefont {Heller}(1981)}]{Heller1981}%
  \BibitemOpen
  \bibfield  {author} {\bibinfo {author} {\bibfnamefont {E.~J.}\ \bibnamefont
  {Heller}},\ }\href@noop {} {\bibfield  {journal} {\bibinfo  {journal} {J.
  Chem. Phys.}\ }\textbf {\bibinfo {volume} {75}},\ \bibinfo {pages} {2923}
  (\bibinfo {year} {1981})}\BibitemShut {NoStop}%
\bibitem [{\citenamefont {Tully}(1990)}]{tul90:1061}%
  \BibitemOpen
  \bibfield  {author} {\bibinfo {author} {\bibfnamefont {J.}~\bibnamefont
  {Tully}},\ }\href {\doibase 10.1063/1.459170} {\bibfield  {journal} {\bibinfo
   {journal} {J. Chem. Phys}\ }\textbf {\bibinfo {volume} {93}},\ \bibinfo
  {pages} {1061} (\bibinfo {year} {1990})}\BibitemShut {NoStop}%
\bibitem [{\citenamefont {Hammes-Schiffer}\ and\ \citenamefont
  {Tully}(1994)}]{ham94:4657}%
  \BibitemOpen
  \bibfield  {author} {\bibinfo {author} {\bibfnamefont {S.}~\bibnamefont
  {Hammes-Schiffer}}\ and\ \bibinfo {author} {\bibfnamefont {J.~C.}\
  \bibnamefont {Tully}},\ }\href {\doibase 10.1063/1.467455} {\bibfield
  {journal} {\bibinfo  {journal} {J. Chem. Phys.}\ }\textbf {\bibinfo {volume}
  {101}},\ \bibinfo {pages} {4657} (\bibinfo {year} {1994})}\BibitemShut
  {NoStop}%
\bibitem [{\citenamefont {Martens}\ and\ \citenamefont
  {Fang}(1997)}]{Martens1997}%
  \BibitemOpen
  \bibfield  {author} {\bibinfo {author} {\bibfnamefont {C.~C.}\ \bibnamefont
  {Martens}}\ and\ \bibinfo {author} {\bibfnamefont {J.-Y.}\ \bibnamefont
  {Fang}},\ }\href@noop {} {\bibfield  {journal} {\bibinfo  {journal} {J. Chem.
  Phys.}\ }\textbf {\bibinfo {volume} {106}},\ \bibinfo {pages} {4918}
  (\bibinfo {year} {1997})}\BibitemShut {NoStop}%
\bibitem [{\citenamefont {Kapral}\ and\ \citenamefont
  {Ciccotti}(1999)}]{Kapral1999}%
  \BibitemOpen
  \bibfield  {author} {\bibinfo {author} {\bibfnamefont {R.}~\bibnamefont
  {Kapral}}\ and\ \bibinfo {author} {\bibfnamefont {G.}~\bibnamefont
  {Ciccotti}},\ }\href@noop {} {\bibfield  {journal} {\bibinfo  {journal} {J.
  Chem. Phys.}\ }\textbf {\bibinfo {volume} {110}},\ \bibinfo {pages} {8919}
  (\bibinfo {year} {1999})}\BibitemShut {NoStop}%
\bibitem [{\citenamefont {Mac~Kernan}, \citenamefont {Ciccotti},\ and\
  \citenamefont {Kapral}(2008)}]{Kernan2008}%
  \BibitemOpen
  \bibfield  {author} {\bibinfo {author} {\bibfnamefont {D.}~\bibnamefont
  {Mac~Kernan}}, \bibinfo {author} {\bibfnamefont {G.}~\bibnamefont
  {Ciccotti}}, \ and\ \bibinfo {author} {\bibfnamefont {R.}~\bibnamefont
  {Kapral}},\ }\href@noop {} {\bibfield  {journal} {\bibinfo  {journal} {J.
  Phys. Chem. B}\ }\textbf {\bibinfo {volume} {112}},\ \bibinfo {pages} {424}
  (\bibinfo {year} {2008})}\BibitemShut {NoStop}%
\bibitem [{\citenamefont {Stock}\ and\ \citenamefont
  {Thoss}(1997)}]{Stock1997}%
  \BibitemOpen
  \bibfield  {author} {\bibinfo {author} {\bibfnamefont {G.}~\bibnamefont
  {Stock}}\ and\ \bibinfo {author} {\bibfnamefont {M.}~\bibnamefont {Thoss}},\
  }\href@noop {} {\bibfield  {journal} {\bibinfo  {journal} {Phys. Rev. Lett.}\
  }\textbf {\bibinfo {volume} {78}},\ \bibinfo {pages} {578} (\bibinfo {year}
  {1997})}\BibitemShut {NoStop}%
\bibitem [{\citenamefont {Thoss}\ and\ \citenamefont
  {Stock}(1999)}]{Thoss1999}%
  \BibitemOpen
  \bibfield  {author} {\bibinfo {author} {\bibfnamefont {M.}~\bibnamefont
  {Thoss}}\ and\ \bibinfo {author} {\bibfnamefont {G.}~\bibnamefont {Stock}},\
  }\href@noop {} {\bibfield  {journal} {\bibinfo  {journal} {Phys. Rev. A}\
  }\textbf {\bibinfo {volume} {59}},\ \bibinfo {pages} {64} (\bibinfo {year}
  {1999})}\BibitemShut {NoStop}%
\bibitem [{\citenamefont {Ben-Nun}, \citenamefont {Quenneville},\ and\
  \citenamefont {Martínez}(2000)}]{Ben2000}%
  \BibitemOpen
  \bibfield  {author} {\bibinfo {author} {\bibfnamefont {M.}~\bibnamefont
  {Ben-Nun}}, \bibinfo {author} {\bibfnamefont {J.}~\bibnamefont
  {Quenneville}}, \ and\ \bibinfo {author} {\bibfnamefont {T.~J.}\ \bibnamefont
  {Martínez}},\ }\href@noop {} {\bibfield  {journal} {\bibinfo  {journal} {J.
  Phys. Chem. A}\ }\textbf {\bibinfo {volume} {104}},\ \bibinfo {pages} {5161}
  (\bibinfo {year} {2000})}\BibitemShut {NoStop}%
\bibitem [{\citenamefont {Curchod}\ and\ \citenamefont
  {Mart\'{i}nez}(2018)}]{Curchod2018}%
  \BibitemOpen
  \bibfield  {author} {\bibinfo {author} {\bibfnamefont {B.~F.~E.}\
  \bibnamefont {Curchod}}\ and\ \bibinfo {author} {\bibfnamefont {T.~J.}\
  \bibnamefont {Mart\'{i}nez}},\ }\href@noop {} {\bibfield  {journal} {\bibinfo
   {journal} {Chem. Rev.}\ }\textbf {\bibinfo {volume} {118}},\ \bibinfo
  {pages} {3305} (\bibinfo {year} {2018})}\BibitemShut {NoStop}%
\bibitem [{\citenamefont {Tavernelli}(2013)}]{Tavernelli2013}%
  \BibitemOpen
  \bibfield  {author} {\bibinfo {author} {\bibfnamefont {I.}~\bibnamefont
  {Tavernelli}},\ }\href@noop {} {\bibfield  {journal} {\bibinfo  {journal}
  {Phys. Rev. A}\ }\textbf {\bibinfo {volume} {87}},\ \bibinfo {pages} {042501}
  (\bibinfo {year} {2013})}\BibitemShut {NoStop}%
\bibitem [{\citenamefont {Curchod}, \citenamefont {Tavernelli},\ and\
  \citenamefont {Rothlisberger}(2011)}]{Curchod2011}%
  \BibitemOpen
  \bibfield  {author} {\bibinfo {author} {\bibfnamefont {B.~F.~E.}\
  \bibnamefont {Curchod}}, \bibinfo {author} {\bibfnamefont {I.}~\bibnamefont
  {Tavernelli}}, \ and\ \bibinfo {author} {\bibfnamefont {U.}~\bibnamefont
  {Rothlisberger}},\ }\href@noop {} {\bibfield  {journal} {\bibinfo  {journal}
  {Phys. Chem. Chem. Phys.}\ }\textbf {\bibinfo {volume} {13}},\ \bibinfo
  {pages} {3231} (\bibinfo {year} {2011})}\BibitemShut {NoStop}%
\bibitem [{\citenamefont {Agostini}\ \emph {et~al.}(2016)\citenamefont
  {Agostini}, \citenamefont {Min}, \citenamefont {Abedi},\ and\ \citenamefont
  {Gross}}]{Agostini2016}%
  \BibitemOpen
  \bibfield  {author} {\bibinfo {author} {\bibfnamefont {F.}~\bibnamefont
  {Agostini}}, \bibinfo {author} {\bibfnamefont {S.~K.}\ \bibnamefont {Min}},
  \bibinfo {author} {\bibfnamefont {A.}~\bibnamefont {Abedi}}, \ and\ \bibinfo
  {author} {\bibfnamefont {E.~K.~U.}\ \bibnamefont {Gross}},\ }\href@noop {}
  {\bibfield  {journal} {\bibinfo  {journal} {J. Chem. Theory Comput.}\
  }\textbf {\bibinfo {volume} {12}},\ \bibinfo {pages} {2127} (\bibinfo {year}
  {2016})}\BibitemShut {NoStop}%
\bibitem [{\citenamefont {Runeson}\ and\ \citenamefont
  {Richardson}(2019)}]{Runeson2019}%
  \BibitemOpen
  \bibfield  {author} {\bibinfo {author} {\bibfnamefont {J.~E.}\ \bibnamefont
  {Runeson}}\ and\ \bibinfo {author} {\bibfnamefont {J.~O.}\ \bibnamefont
  {Richardson}},\ }\href@noop {} {\bibfield  {journal} {\bibinfo  {journal} {J.
  Chem. Phys.}\ }\textbf {\bibinfo {volume} {151}},\ \bibinfo {pages} {044119}
  (\bibinfo {year} {2019})}\BibitemShut {NoStop}%
\bibitem [{\citenamefont {Runeson}\ and\ \citenamefont
  {Richardson}(2020)}]{Runeson2020}%
  \BibitemOpen
  \bibfield  {author} {\bibinfo {author} {\bibfnamefont {J.~E.}\ \bibnamefont
  {Runeson}}\ and\ \bibinfo {author} {\bibfnamefont {J.~O.}\ \bibnamefont
  {Richardson}},\ }\href {\doibase 10.1063/1.5143412} {\bibfield  {journal}
  {\bibinfo  {journal} {J. Chem. Phys.}\ }\textbf {\bibinfo {volume} {152}},\
  \bibinfo {pages} {084110} (\bibinfo {year} {2020})},\ \Eprint
  {http://arxiv.org/abs/https://doi.org/10.1063/1.5143412}
  {https://doi.org/10.1063/1.5143412} \BibitemShut {NoStop}%
\bibitem [{\citenamefont {Lang}, \citenamefont {Vendrell},\ and\ \citenamefont
  {Hauke}(2021)}]{lan21:024111}%
  \BibitemOpen
  \bibfield  {author} {\bibinfo {author} {\bibfnamefont {H.}~\bibnamefont
  {Lang}}, \bibinfo {author} {\bibfnamefont {O.}~\bibnamefont {Vendrell}}, \
  and\ \bibinfo {author} {\bibfnamefont {P.}~\bibnamefont {Hauke}},\ }\href
  {\doibase 10.1063/5.0054696} {\bibfield  {journal} {\bibinfo  {journal} {J.
  Chem. Phys.}\ }\textbf {\bibinfo {volume} {155}},\ \bibinfo {pages} {024111}
  (\bibinfo {year} {2021})}\BibitemShut {NoStop}%
\bibitem [{\citenamefont {Cotton}\ and\ \citenamefont
  {Miller}(2015{\natexlab{a}})}]{Cotton2015}%
  \BibitemOpen
  \bibfield  {author} {\bibinfo {author} {\bibfnamefont {S.~J.}\ \bibnamefont
  {Cotton}}\ and\ \bibinfo {author} {\bibfnamefont {W.~H.}\ \bibnamefont
  {Miller}},\ }\href@noop {} {\bibfield  {journal} {\bibinfo  {journal} {J.
  Phys. Chem. A}\ }\textbf {\bibinfo {volume} {119}},\ \bibinfo {pages} {12138}
  (\bibinfo {year} {2015}{\natexlab{a}})}\BibitemShut {NoStop}%
\bibitem [{\citenamefont {Cotton}\ and\ \citenamefont
  {Miller}(2013)}]{Cotton2013}%
  \BibitemOpen
  \bibfield  {author} {\bibinfo {author} {\bibfnamefont {S.~J.}\ \bibnamefont
  {Cotton}}\ and\ \bibinfo {author} {\bibfnamefont {W.~H.}\ \bibnamefont
  {Miller}},\ }\href@noop {} {\bibfield  {journal} {\bibinfo  {journal} {J.
  Chem. Phys.}\ }\textbf {\bibinfo {volume} {139}},\ \bibinfo {pages} {234112}
  (\bibinfo {year} {2013})}\BibitemShut {NoStop}%
\bibitem [{\citenamefont {Cotton}\ and\ \citenamefont
  {Miller}(2015{\natexlab{b}})}]{cot15:12138}%
  \BibitemOpen
  \bibfield  {author} {\bibinfo {author} {\bibfnamefont {S.~J.}\ \bibnamefont
  {Cotton}}\ and\ \bibinfo {author} {\bibfnamefont {W.~H.}\ \bibnamefont
  {Miller}},\ }\href {\doibase 10.1021/acs.jpca.5b05906} {\bibfield  {journal}
  {\bibinfo  {journal} {J. Phys. Chem. A}\ }\textbf {\bibinfo {volume} {119}},\
  \bibinfo {pages} {12138} (\bibinfo {year} {2015}{\natexlab{b}})}\BibitemShut
  {NoStop}%
\bibitem [{\citenamefont {Liang}\ \emph {et~al.}(2018)\citenamefont {Liang},
  \citenamefont {Cotton}, \citenamefont {Binder}, \citenamefont {Hegger},
  \citenamefont {Burghardt},\ and\ \citenamefont {Miller}}]{Liang2018}%
  \BibitemOpen
  \bibfield  {author} {\bibinfo {author} {\bibfnamefont {R.}~\bibnamefont
  {Liang}}, \bibinfo {author} {\bibfnamefont {S.~J.}\ \bibnamefont {Cotton}},
  \bibinfo {author} {\bibfnamefont {R.}~\bibnamefont {Binder}}, \bibinfo
  {author} {\bibfnamefont {R.}~\bibnamefont {Hegger}}, \bibinfo {author}
  {\bibfnamefont {I.}~\bibnamefont {Burghardt}}, \ and\ \bibinfo {author}
  {\bibfnamefont {W.~H.}\ \bibnamefont {Miller}},\ }\href {\doibase
  10.1063/1.5037815} {\bibfield  {journal} {\bibinfo  {journal} {J. Chem.
  Phys.}\ }\textbf {\bibinfo {volume} {149}},\ \bibinfo {pages} {044101}
  (\bibinfo {year} {2018})}\BibitemShut {NoStop}%
\bibitem [{\citenamefont {Gao}\ \emph {et~al.}(2020)\citenamefont {Gao},
  \citenamefont {Saller}, \citenamefont {Liu}, \citenamefont {Kelly},
  \citenamefont {Richardson},\ and\ \citenamefont {Geva}}]{Gao2020}%
  \BibitemOpen
  \bibfield  {author} {\bibinfo {author} {\bibfnamefont {X.}~\bibnamefont
  {Gao}}, \bibinfo {author} {\bibfnamefont {M.~A.~C.}\ \bibnamefont {Saller}},
  \bibinfo {author} {\bibfnamefont {Y.}~\bibnamefont {Liu}}, \bibinfo {author}
  {\bibfnamefont {A.}~\bibnamefont {Kelly}}, \bibinfo {author} {\bibfnamefont
  {J.~O.}\ \bibnamefont {Richardson}}, \ and\ \bibinfo {author} {\bibfnamefont
  {E.}~\bibnamefont {Geva}},\ }\href {\doibase 10.1021/acs.jctc.9b01267}
  {\bibfield  {journal} {\bibinfo  {journal} {J. Chem. Theory Comput.}\
  }\textbf {\bibinfo {volume} {16}},\ \bibinfo {pages} {2883} (\bibinfo {year}
  {2020})},\ \bibinfo {note} {pMID: 32227993}\BibitemShut {NoStop}%
\bibitem [{\citenamefont {Miller}(1976)}]{Miller1976}%
  \BibitemOpen
  \bibfield  {author} {\bibinfo {author} {\bibfnamefont {W.~H.}\ \bibnamefont
  {Miller}},\ }\href {\doibase 10.1063/1.432590} {\bibfield  {journal}
  {\bibinfo  {journal} {J. Chem. Phys.}\ }\textbf {\bibinfo {volume} {64}},\
  \bibinfo {pages} {2880} (\bibinfo {year} {1976})}\BibitemShut {NoStop}%
\bibitem [{\citenamefont {Miller}\ and\ \citenamefont
  {McCurdy}(1978)}]{Miller1978}%
  \BibitemOpen
  \bibfield  {author} {\bibinfo {author} {\bibfnamefont {W.~H.}\ \bibnamefont
  {Miller}}\ and\ \bibinfo {author} {\bibfnamefont {C.~W.}\ \bibnamefont
  {McCurdy}},\ }\href {\doibase 10.1063/1.436463} {\bibfield  {journal}
  {\bibinfo  {journal} {J. Chem. Phys.}\ }\textbf {\bibinfo {volume} {69}},\
  \bibinfo {pages} {5163} (\bibinfo {year} {1978})}\BibitemShut {NoStop}%
\bibitem [{\citenamefont {Jordan}\ and\ \citenamefont
  {Wigner}(1928)}]{jor28:631}%
  \BibitemOpen
  \bibfield  {author} {\bibinfo {author} {\bibfnamefont {P.}~\bibnamefont
  {Jordan}}\ and\ \bibinfo {author} {\bibfnamefont {E.}~\bibnamefont
  {Wigner}},\ }\href {\doibase 10.1007/bf01331938} {\bibfield  {journal}
  {\bibinfo  {journal} {Zeitschrift f{\"u}r Physik}\ }\textbf {\bibinfo
  {volume} {47}},\ \bibinfo {pages} {631} (\bibinfo {year} {1928})}\BibitemShut
  {NoStop}%
\bibitem [{\citenamefont {Li}\ and\ \citenamefont {Miller}(2012)}]{Li2012}%
  \BibitemOpen
  \bibfield  {author} {\bibinfo {author} {\bibfnamefont {B.}~\bibnamefont
  {Li}}\ and\ \bibinfo {author} {\bibfnamefont {W.~H.}\ \bibnamefont
  {Miller}},\ }\href@noop {} {\bibfield  {journal} {\bibinfo  {journal} {J.
  Chem. Phys.}\ }\textbf {\bibinfo {volume} {137}},\ \bibinfo {pages} {154107}
  (\bibinfo {year} {2012})}\BibitemShut {NoStop}%
\bibitem [{\citenamefont {Li}\ \emph {et~al.}(2013)\citenamefont {Li},
  \citenamefont {Levy}, \citenamefont {Swenson}, \citenamefont {Rabani},\ and\
  \citenamefont {Miller}}]{Li2013}%
  \BibitemOpen
  \bibfield  {author} {\bibinfo {author} {\bibfnamefont {B.}~\bibnamefont
  {Li}}, \bibinfo {author} {\bibfnamefont {T.~J.}\ \bibnamefont {Levy}},
  \bibinfo {author} {\bibfnamefont {D.~W.~H.}\ \bibnamefont {Swenson}},
  \bibinfo {author} {\bibfnamefont {E.}~\bibnamefont {Rabani}}, \ and\ \bibinfo
  {author} {\bibfnamefont {W.~H.}\ \bibnamefont {Miller}},\ }\href@noop {}
  {\bibfield  {journal} {\bibinfo  {journal} {J. Chem. Phys.}\ }\textbf
  {\bibinfo {volume} {138}},\ \bibinfo {pages} {104110} (\bibinfo {year}
  {2013})}\BibitemShut {NoStop}%
\bibitem [{\citenamefont {Li}\ \emph {et~al.}(2014)\citenamefont {Li},
  \citenamefont {Miller}, \citenamefont {Levy},\ and\ \citenamefont
  {Rabani}}]{Li2014}%
  \BibitemOpen
  \bibfield  {author} {\bibinfo {author} {\bibfnamefont {B.}~\bibnamefont
  {Li}}, \bibinfo {author} {\bibfnamefont {W.~H.}\ \bibnamefont {Miller}},
  \bibinfo {author} {\bibfnamefont {T.~J.}\ \bibnamefont {Levy}}, \ and\
  \bibinfo {author} {\bibfnamefont {E.}~\bibnamefont {Rabani}},\ }\href@noop {}
  {\bibfield  {journal} {\bibinfo  {journal} {J. Chem. Phys.}\ }\textbf
  {\bibinfo {volume} {140}},\ \bibinfo {pages} {204106} (\bibinfo {year}
  {2014})}\BibitemShut {NoStop}%
\bibitem [{\citenamefont {Levy}\ \emph {et~al.}(2019)\citenamefont {Levy},
  \citenamefont {Dou}, \citenamefont {Rabani},\ and\ \citenamefont
  {Limmer}}]{Levy2019}%
  \BibitemOpen
  \bibfield  {author} {\bibinfo {author} {\bibfnamefont {A.}~\bibnamefont
  {Levy}}, \bibinfo {author} {\bibfnamefont {W.}~\bibnamefont {Dou}}, \bibinfo
  {author} {\bibfnamefont {E.}~\bibnamefont {Rabani}}, \ and\ \bibinfo {author}
  {\bibfnamefont {D.~T.}\ \bibnamefont {Limmer}},\ }\href@noop {} {\bibfield
  {journal} {\bibinfo  {journal} {J. Chem. Phys.}\ }\textbf {\bibinfo {volume}
  {150}},\ \bibinfo {pages} {234112} (\bibinfo {year} {2019})}\BibitemShut
  {NoStop}%
\bibitem [{\citenamefont {Zwanziger}, \citenamefont {Grant},\ and\
  \citenamefont {Ezra}(1986)}]{Zwanziger1986}%
  \BibitemOpen
  \bibfield  {author} {\bibinfo {author} {\bibfnamefont {J.~W.}\ \bibnamefont
  {Zwanziger}}, \bibinfo {author} {\bibfnamefont {E.~R.}\ \bibnamefont
  {Grant}}, \ and\ \bibinfo {author} {\bibfnamefont {G.~S.}\ \bibnamefont
  {Ezra}},\ }\href {\doibase 10.1063/1.451153} {\bibfield  {journal} {\bibinfo
  {journal} {J. Chem. Phys.}\ }\textbf {\bibinfo {volume} {85}},\ \bibinfo
  {pages} {2089} (\bibinfo {year} {1986})}\BibitemShut {NoStop}%
\bibitem [{\citenamefont {Van~Voorhis}\ and\ \citenamefont
  {Reichman}(2004)}]{Voorhis2004}%
  \BibitemOpen
  \bibfield  {author} {\bibinfo {author} {\bibfnamefont {T.}~\bibnamefont
  {Van~Voorhis}}\ and\ \bibinfo {author} {\bibfnamefont {D.~R.}\ \bibnamefont
  {Reichman}},\ }\href {\doibase 10.1063/1.1630963} {\bibfield  {journal}
  {\bibinfo  {journal} {J. Chem. Phys.}\ }\textbf {\bibinfo {volume} {120}},\
  \bibinfo {pages} {579} (\bibinfo {year} {2004})}\BibitemShut {NoStop}%
\bibitem [{\citenamefont {Swenson}\ \emph {et~al.}(2011)\citenamefont
  {Swenson}, \citenamefont {Levy}, \citenamefont {Cohen}, \citenamefont
  {Rabani},\ and\ \citenamefont {Miller}}]{Swenson2011}%
  \BibitemOpen
  \bibfield  {author} {\bibinfo {author} {\bibfnamefont {D.~W.~H.}\
  \bibnamefont {Swenson}}, \bibinfo {author} {\bibfnamefont {T.}~\bibnamefont
  {Levy}}, \bibinfo {author} {\bibfnamefont {G.}~\bibnamefont {Cohen}},
  \bibinfo {author} {\bibfnamefont {E.}~\bibnamefont {Rabani}}, \ and\ \bibinfo
  {author} {\bibfnamefont {W.~H.}\ \bibnamefont {Miller}},\ }\href@noop {}
  {\bibfield  {journal} {\bibinfo  {journal} {J. Chem. Phys.}\ }\textbf
  {\bibinfo {volume} {134}},\ \bibinfo {pages} {164103} (\bibinfo {year}
  {2011})}\BibitemShut {NoStop}%
\bibitem [{\citenamefont {Liu}(2016)}]{Liu2016}%
  \BibitemOpen
  \bibfield  {author} {\bibinfo {author} {\bibfnamefont {J.}~\bibnamefont
  {Liu}},\ }\href {\doibase 10.1063/1.4967815} {\bibfield  {journal} {\bibinfo
  {journal} {J. Chem. Phys.}\ }\textbf {\bibinfo {volume} {145}},\ \bibinfo
  {pages} {204105} (\bibinfo {year} {2016})},\ \Eprint
  {http://arxiv.org/abs/https://doi.org/10.1063/1.4967815}
  {https://doi.org/10.1063/1.4967815} \BibitemShut {NoStop}%
\bibitem [{\citenamefont {Cotton}\ and\ \citenamefont
  {Miller}(2016)}]{cot16:983}%
  \BibitemOpen
  \bibfield  {author} {\bibinfo {author} {\bibfnamefont {S.~J.}\ \bibnamefont
  {Cotton}}\ and\ \bibinfo {author} {\bibfnamefont {W.~H.}\ \bibnamefont
  {Miller}},\ }\href {\doibase 10.1021/acs.jctc.5b01178} {\bibfield  {journal}
  {\bibinfo  {journal} {J. Chem. Theory Comput.}\ }\textbf {\bibinfo {volume}
  {12}},\ \bibinfo {pages} {983} (\bibinfo {year} {2016})}\BibitemShut
  {NoStop}%
\bibitem [{\citenamefont {Remacle}\ and\ \citenamefont
  {Levine}(2000)}]{Remacle2000}%
  \BibitemOpen
  \bibfield  {author} {\bibinfo {author} {\bibfnamefont {F.}~\bibnamefont
  {Remacle}}\ and\ \bibinfo {author} {\bibfnamefont {R.~D.}\ \bibnamefont
  {Levine}},\ }\href@noop {} {\bibfield  {journal} {\bibinfo  {journal} {J.
  Chem. Phys.}\ }\textbf {\bibinfo {volume} {113}},\ \bibinfo {pages} {4515}
  (\bibinfo {year} {2000})}\BibitemShut {NoStop}%
\bibitem [{\citenamefont {Tannor}(2007)}]{tannor-book}%
  \BibitemOpen
  \bibfield  {author} {\bibinfo {author} {\bibfnamefont {D.~J.}\ \bibnamefont
  {Tannor}},\ }\href@noop {} {\emph {\bibinfo {title} {Introduction to quantum
  mechanics: a time-dependent perspective}}},\ edited by\ \bibinfo {editor}
  {\bibfnamefont {L.~A.}\ \bibnamefont {Young}}\ (\bibinfo  {publisher}
  {University Science Books},\ \bibinfo {year} {2007})\BibitemShut {NoStop}%
\bibitem [{\citenamefont {{\.{I}}mre}\ \emph {et~al.}(1967)\citenamefont
  {{\.{I}}mre}, \citenamefont {{{\"{O}}}zizmir}, \citenamefont {Rosenbaum},\
  and\ \citenamefont {Zweifel}}]{imr67:1097}%
  \BibitemOpen
  \bibfield  {author} {\bibinfo {author} {\bibfnamefont {K.}~\bibnamefont
  {{\.{I}}mre}}, \bibinfo {author} {\bibfnamefont {E.}~\bibnamefont
  {{{\"{O}}}zizmir}}, \bibinfo {author} {\bibfnamefont {M.}~\bibnamefont
  {Rosenbaum}}, \ and\ \bibinfo {author} {\bibfnamefont {P.~F.}\ \bibnamefont
  {Zweifel}},\ }\href {\doibase 10.1063/1.1705323} {\bibfield  {journal}
  {\bibinfo  {journal} {J. Math. Phys.}\ }\textbf {\bibinfo {volume} {8}},\
  \bibinfo {pages} {1097} (\bibinfo {year} {1967})}\BibitemShut {NoStop}%
\bibitem [{\citenamefont {Meyer}, \citenamefont {Manthe},\ and\ \citenamefont
  {Cederbaum}(1990)}]{Meyer1990}%
  \BibitemOpen
  \bibfield  {author} {\bibinfo {author} {\bibfnamefont {H.-D.}\ \bibnamefont
  {Meyer}}, \bibinfo {author} {\bibfnamefont {U.}~\bibnamefont {Manthe}}, \
  and\ \bibinfo {author} {\bibfnamefont {L.}~\bibnamefont {Cederbaum}},\
  }\href@noop {} {\bibfield  {journal} {\bibinfo  {journal} {Chem. Phys.
  Lett.}\ }\textbf {\bibinfo {volume} {165}},\ \bibinfo {pages} {73} (\bibinfo
  {year} {1990})}\BibitemShut {NoStop}%
\bibitem [{\citenamefont {Beck}\ \emph {et~al.}(2000)\citenamefont {Beck},
  \citenamefont {J\"ackle}, \citenamefont {Worth},\ and\ \citenamefont
  {Meyer}}]{bec00:1}%
  \BibitemOpen
  \bibfield  {author} {\bibinfo {author} {\bibfnamefont {M.~H.}\ \bibnamefont
  {Beck}}, \bibinfo {author} {\bibfnamefont {A.}~\bibnamefont {J\"ackle}},
  \bibinfo {author} {\bibfnamefont {G.~A.}\ \bibnamefont {Worth}}, \ and\
  \bibinfo {author} {\bibfnamefont {H.-D.}\ \bibnamefont {Meyer}},\ }\href
  {\doibase 10.1016/S0370-1573(99)00047-2} {\bibfield  {journal} {\bibinfo
  {journal} {Phys. Rep.}\ }\textbf {\bibinfo {volume} {324}},\ \bibinfo {pages}
  {1} (\bibinfo {year} {2000})}\BibitemShut {NoStop}%
\bibitem [{\citenamefont {Worth}\ \emph {et~al.}()\citenamefont {Worth},
  \citenamefont {Beck}, \citenamefont {J{\"a}ckle}, \citenamefont {Vendrell},\
  and\ \citenamefont {Meyer}}]{mctdh:MLpackage}%
  \BibitemOpen
  \bibfield  {author} {\bibinfo {author} {\bibfnamefont {G.~A.}\ \bibnamefont
  {Worth}}, \bibinfo {author} {\bibfnamefont {M.~H.}\ \bibnamefont {Beck}},
  \bibinfo {author} {\bibfnamefont {A.}~\bibnamefont {J{\"a}ckle}}, \bibinfo
  {author} {\bibfnamefont {O.}~\bibnamefont {Vendrell}}, \ and\ \bibinfo
  {author} {\bibfnamefont {H.-D.}\ \bibnamefont {Meyer}},\ }\href@noop {}
  {}\bibinfo {howpublished} {The {MCTDH} {P}ackage, {V}ersion 8.2, (2000).
  H.-D. Meyer, {V}ersion 8.3 (2002), {V}ersion 8.4 (2007). O. Vendrell and
  H.-D. Meyer {V}ersion 8.5 (2013). {V}ersion 8.5 contains the ML-MCTDH
  algorithm. Current versions: 8.4.18 and 8.5.11 (2019). {S}ee
  http://mctdh.uni-hd.de/}\BibitemShut {NoStop}%
\bibitem [{\citenamefont {Wang}\ and\ \citenamefont
  {Thoss}(2009)}]{wan09:024114}%
  \BibitemOpen
  \bibfield  {author} {\bibinfo {author} {\bibfnamefont {H.}~\bibnamefont
  {Wang}}\ and\ \bibinfo {author} {\bibfnamefont {M.}~\bibnamefont {Thoss}},\
  }\href {\doibase 10.1063/1.3173823} {\bibfield  {journal} {\bibinfo
  {journal} {J. Chem. Phys.}\ }\textbf {\bibinfo {volume} {131}},\ \bibinfo
  {pages} {024114} (\bibinfo {year} {2009})}\BibitemShut {NoStop}%
\bibitem [{\citenamefont {Manthe}\ and\ \citenamefont
  {Weike}(2017)}]{man17:064117}%
  \BibitemOpen
  \bibfield  {author} {\bibinfo {author} {\bibfnamefont {U.}~\bibnamefont
  {Manthe}}\ and\ \bibinfo {author} {\bibfnamefont {T.}~\bibnamefont {Weike}},\
  }\href {\doibase 10.1063/1.4975662} {\bibfield  {journal} {\bibinfo
  {journal} {J. Chem. Phys.}\ }\textbf {\bibinfo {volume} {146}},\ \bibinfo
  {pages} {064117} (\bibinfo {year} {2017})}\BibitemShut {NoStop}%
\bibitem [{\citenamefont {Sasmal}\ and\ \citenamefont
  {Vendrell}(2020)}]{sas2020:154110}%
  \BibitemOpen
  \bibfield  {author} {\bibinfo {author} {\bibfnamefont {S.}~\bibnamefont
  {Sasmal}}\ and\ \bibinfo {author} {\bibfnamefont {O.}~\bibnamefont
  {Vendrell}},\ }\href {\doibase 10.1063/5.0028116} {\bibfield  {journal}
  {\bibinfo  {journal} {J. Chem. Phys.}\ }\textbf {\bibinfo {volume} {153}},\
  \bibinfo {pages} {154110} (\bibinfo {year} {2020})}\BibitemShut {NoStop}%
\bibitem [{\citenamefont {Jordan}\ and\ \citenamefont
  {Wigner}(1993)}]{Jordan1993}%
  \BibitemOpen
  \bibfield  {author} {\bibinfo {author} {\bibfnamefont {P.}~\bibnamefont
  {Jordan}}\ and\ \bibinfo {author} {\bibfnamefont {E.~P.}\ \bibnamefont
  {Wigner}},\ }\enquote {\bibinfo {title} {{\"U}ber das paulische
  {\"a}quivalenzverbot},}\ in\ \href@noop {} {\emph {\bibinfo {booktitle} {The
  Collected Works of Eugene Paul Wigner: Part A: The Scientific Papers}}},\
  \bibinfo {editor} {edited by\ \bibinfo {editor} {\bibfnamefont {A.~S.}\
  \bibnamefont {Wightman}}}\ (\bibinfo  {publisher} {Springer Berlin
  Heidelberg},\ \bibinfo {year} {1993})\ pp.\ \bibinfo {pages}
  {109--129}\BibitemShut {NoStop}%
\bibitem [{\citenamefont {Tamura}\ \emph {et~al.}(2015)\citenamefont {Tamura},
  \citenamefont {Huix-Rotllant}, \citenamefont {Burghardt}, \citenamefont
  {Olivier},\ and\ \citenamefont {Beljonne}}]{tam15:107401}%
  \BibitemOpen
  \bibfield  {author} {\bibinfo {author} {\bibfnamefont {H.}~\bibnamefont
  {Tamura}}, \bibinfo {author} {\bibfnamefont {M.}~\bibnamefont
  {Huix-Rotllant}}, \bibinfo {author} {\bibfnamefont {I.}~\bibnamefont
  {Burghardt}}, \bibinfo {author} {\bibfnamefont {Y.}~\bibnamefont {Olivier}},
  \ and\ \bibinfo {author} {\bibfnamefont {D.}~\bibnamefont {Beljonne}},\
  }\href {\doibase 10.1103/physrevlett.115.107401} {\bibfield  {journal}
  {\bibinfo  {journal} {Phys. Rev. Lett.}\ }\textbf {\bibinfo {volume} {115}},\
  \bibinfo {pages} {107401} (\bibinfo {year} {2015})}\BibitemShut {NoStop}%
\bibitem [{\citenamefont {Ford}\ and\ \citenamefont
  {Andrews}(2014)}]{Ford2014}%
  \BibitemOpen
  \bibfield  {author} {\bibinfo {author} {\bibfnamefont {J.~S.}\ \bibnamefont
  {Ford}}\ and\ \bibinfo {author} {\bibfnamefont {D.~L.}\ \bibnamefont
  {Andrews}},\ }\href@noop {} {\bibfield  {journal} {\bibinfo  {journal} {Chem.
  Phys. Lett.}\ }\textbf {\bibinfo {volume} {591}},\ \bibinfo {pages} {88 }
  (\bibinfo {year} {2014})}\BibitemShut {NoStop}%
\bibitem [{\citenamefont {Tomasi}\ and\ \citenamefont
  {Kassal}(2020)}]{tom20:2348}%
  \BibitemOpen
  \bibfield  {author} {\bibinfo {author} {\bibfnamefont {S.}~\bibnamefont
  {Tomasi}}\ and\ \bibinfo {author} {\bibfnamefont {I.}~\bibnamefont
  {Kassal}},\ }\href {\doibase 10.1021/acs.jpclett.9b03490} {\bibfield
  {journal} {\bibinfo  {journal} {J. Phys. Chem. Lett.}\ }\textbf {\bibinfo
  {volume} {11}},\ \bibinfo {pages} {2348} (\bibinfo {year}
  {2020})}\BibitemShut {NoStop}%
\bibitem [{\citenamefont {Meyer}\ and\ \citenamefont
  {Miller}(1980{\natexlab{b}})}]{Meyer1980}%
  \BibitemOpen
  \bibfield  {author} {\bibinfo {author} {\bibfnamefont {H.}~\bibnamefont
  {Meyer}}\ and\ \bibinfo {author} {\bibfnamefont {W.~H.}\ \bibnamefont
  {Miller}},\ }\href {\doibase 10.1063/1.439462} {\bibfield  {journal}
  {\bibinfo  {journal} {J. Chem. Phys.}\ }\textbf {\bibinfo {volume} {72}},\
  \bibinfo {pages} {2272} (\bibinfo {year} {1980}{\natexlab{b}})},\ \Eprint
  {http://arxiv.org/abs/https://doi.org/10.1063/1.439462}
  {https://doi.org/10.1063/1.439462} \BibitemShut {NoStop}%
\end{thebibliography}
%

\end{document}


\title{A bosonic perspective on the classical mapping of fermionic quantum dynamics:\\
Supporting Information}

\author{Jing Sun} 
\author{Sudip Sasmal}%
\author{Oriol Vendrell}%
\affiliation{
    Theoretische Chemie, Physikalisch-Chemisches Institut, Universität
    Heidelberg, Germany.
}
\maketitle
\beginsupplement


\begin{table}
\begin{tabular}{c c c c c c c c c c c c}
 \hline
 \\
 dimer(MO) & $h_{h_{I},h_{I}}$ & $h_{l_{I},l_{I}}$ & $h_{h_{II},h_{II}}$ & $h_{l_{II},l_{II}}$ & $V^{h_{II}h_{II}}_{h_{I}h_{I}}$ & $V^{l_{II}l_{II}}_{h_{I}h_{I}}$ & $V^{h_{II}h_{II}}_{l_{I}l_{I}}$ & $V^{l_{II}l_{II}}_{l_{I}l_{I}}$ & $V^{h_{II}l_{II}}_{h_{I}l_{I}}$ & & \\
 \\
  & -0.9382 & -0.6341 & -0.9382 & -0.6341 & 0.05304 & 0.05296 & 0.05296 & 0.05288 & 0.0002555 & & \\
 \\
 trimer & $h_{h_{I},h_{I}}$ & $h_{l_{I},l_{I}}$ & $h_{h_{II},h_{II}}$ & $h_{l_{II},l_{II}}$ & $h_{h_{III},h_{III}}$ & $h_{l_{III},l_{III}}$ & $V^{h_{II}h_{II}}_{h_{I}h_{I}}$ & $V^{l_{II}l_{II}}_{h_{I}h_{I}}$ & $V^{h_{II}h_{II}}_{l_{I}l_{I}}$ & $V^{l_{II}l_{II}}_{l_{I}l_{I}}$ & $V^{h_{III}h_{III}}_{h_{I}h_{I}}$\\
 \\
  & -0.9911 & -0.6870 & -1.044 & -0.7398 & -0.9911 & -0.6870 & 0.05304 & 0.05296 & 0.05296 & 0.05288 & 0.02647\\
  \\
    & $V^{l_{III}l_{III}}_{h_{I}h_{I}}$&
    $V^{h_{III}h_{III}}_{l_{I}l_{I}}$ & $V^{l_{III}l_{III}}_{l_{I}l_{I}}$ & $V^{h_{III}h_{III}}_{h_{II}h_{II}}$ & $V^{l_{III}l_{III}}_{h_{II}h_{II}}$ & $V^{h_{III}h_{III}}_{l_{II}l_{II}}$ & $V^{l_{III}l_{III}}_{l_{II}l_{II}}$ & & & & \\
    \\
     & 0.02646 & 0.02646 & 0.02645 & 0.05304 & 0.05296 & 0.05296 & 0.05288 & & & & \\
     \\
 $\theta=0$ & $V^{h_{II}l_{II}}_{h_{I}l_{I}}$ & & & & & & & & & & \\
 \\
  & 0.0002555 & & & & & & & & & & \\
  \\
 $\theta=\frac{\pi}{6}$ & $V^{h_{II}l_{II}}_{h_{I}l_{I}}$ & $V^{h_{III}l_{III}}_{h_{II}l_{II}}$ & & & & & & & & & \\
 \\
  & 0.0002212 & 0.0001278 & & & & & & & & & \\
  \\
 $\theta=\frac{\pi}{4}$ & $V^{h_{II}l_{II}}_{h_{I}l_{I}}$ & $V^{h_{III}l_{III}}_{h_{II}l_{II}}$ & & & & & & & & & \\
 \\
  & 0.0001806 & 0.0001806 & & & & & & & & & \\
  \\
 tetramer & $h_{h_{I},h_{I}}$ & $h_{l_{I},l_{I}}$ & $h_{h_{II},h_{II}}$ & $h_{l_{II},l_{II}}$ & $h_{h_{III},h_{III}}$ & $h_{l_{III},l_{III}}$ & $h_{h_{IV},h_{IV}}$ & $h_{l_{IV},l_{IV}}$ & $V^{h_{II}h_{II}}_{h_{I}h_{I}}$ & $V^{l_{II}l_{II}}_{h_{I}h_{I}}$ & $V^{h_{II}h_{II}}_{l_{I}l_{I}}$\\
 \\
  & -1.074 & -0.7703 & -1.126 & -0.8225 & -1.126 & -0.8225 & -1.074 & -0.7703 & 0.04734 & 0.04730 & 0.04730\\
  \\
   & $V^{l_{II}l_{II}}_{l_{I}l_{I}}$ & $V^{h_{III}h_{III}}_{h_{I}h_{I}}$ & $V^{l_{III}l_{III}}_{h_{I}h_{I}}$ & $V^{h_{III}h_{III}}_{l_{I}l_{I}}$ & $V^{l_{III}l_{III}}_{l_{I}l_{I}}$ & $V^{h_{IV}h_{IV}}_{h_{I}h_{I}}$ & $V^{l_{IV}l_{IV}}_{h_{I}h_{I}}$ & $V^{h_{IV}h_{IV}}_{l_{I}l_{I}}$ & $V^{l_{IV}l_{IV}}_{l_{I}l_{I}}$ & $V^{h_{III}h_{III}}_{h_{II}h_{II}}$ & $V^{l_{III}l_{III}}_{h_{II}h_{II}}$ \\
   \\
  & 0.04725 & 0.04734 & 0.04730 & 0.04730 & 0.04725 & 0.02647 & 0.02646 & 0.02646 & 0.02645 & 0.05256 & 0.05252\\
  \\
 & $V^{h_{III}h_{III}}_{l_{II}l_{II}}$ & $V^{l_{III}l_{III}}_{l_{II}l_{II}}$ & $V^{h_{IV}h_{IV}}_{h_{II}h_{II}}$ & $V^{l_{IV}l_{IV}}_{h_{II}h_{II}}$ & $V^{h_{IV}h_{IV}}_{l_{II}l_{II}}$ & $V^{l_{IV}l_{IV}}_{l_{II}l_{II}}$ & $V^{h_{IV}h_{IV}}_{h_{III}h_{III}}$ & $V^{l_{IV}l_{IV}}_{h_{III}h_{III}}$ & $V^{h_{IV}h_{IV}}_{l_{III}l_{III}}$ & $V^{l_{IV}l_{IV}}_{l_{III}l_{III}}$ & \\
 \\
   & 0.05252 & 0.05249 & 0.04734 & 0.04730 & 0.04730 & 0.04725 & 0.04734 & 0.04730 & 0.04730 & 0.04725 &\\
   \\
 (cons) & $V^{h_{II}l_{II}}_{h_{I}l_{I}}$ & $V^{h_{III}l_{III}}_{h_{I}l_{I}}$ & $V^{h_{II}l_{II}}_{h_{IV}l_{IV}}$ & $V^{h_{III}l_{III}}_{h_{IV}l_{IV}}$ & & & & & & &\\
 \\
  & 0.0001809 & 0.0001809 & 0.0001809 & 0.0001809  & & & & & & &\\
  \\
 (des) & $V^{h_{II}l_{II}}_{h_{I}l_{I}}$ & $V^{h_{III}l_{III}}_{h_{I}l_{I}}$ & $V^{h_{II}l_{II}}_{h_{IV}l_{IV}}$ & $V^{h_{III}l_{III}}_{h_{IV}l_{IV}}$  & & & & & & &\\
 \\
  & 0.0001809 & 0.0001809 & -0.0001809 & 0.0001809  & & & & & & &\\
 \hline
\end{tabular}
\caption{\label{Tab:Int_ethy} Unique, non-zero integrals in the ethylene dimer, trimer and tetramer clusters. 
%
%
All integrals correspond to localized Hartree-Fock molecular orbitals obtained using a minimal basis.
The indices $h_I$/$l_I$, e.g., represent the HOMO/LUMO orbitals of the $I$-th ethylene monomer. All energies are
in Hartree units.}

\end{table}